\documentclass[prb,showpacs,twocolumn,superscriptaddress]{revtex4}

\usepackage{amsmath,graphicx,latexsym,bm}

\begin{document}

\title{Electrostatic stability of insulating surfaces: Theory and applications}

\author{Massimiliano Stengel}
\affiliation{Institut de Ci\`encia de Materials de Barcelona 
(ICMAB-CSIC), Campus UAB, 08193 Bellaterra, Spain}

\date{\today}

\begin{abstract}
We analyze the electrostatic stability of insulating surfaces in the
framework of the bulk modern theory of polarization. We show that heuristic
arguments based on a fully ionic limit find formal justification at the 
microscopic level, even in solids where the bonding has a mixed ionic/covalent
character. Based on these arguments, we propose simple criteria to construct
arbitrary non-polar terminations of a given bulk crystal. We illustrate 
our ideas by performing model calculations of several LaAlO$_3$ and 
SrTiO$_3$ surfaces. We find, in the case of ideal LaAlO$_3$ surfaces, that 
cleavage along a higher-index $(n10)$ direction is energetically favorable 
compared to the polar (100) orientation. In the presence of external
adsorbates or defects the picture can change dramatically, as we
demonstrate in the case of H$_2$O/LaAlO$_3$(100).
\end{abstract}


\maketitle

\section{Introduction}

A uniformly charged plane separating two semi-infinite regions of space 
yields a divergent electrostatic energy; for this reason, \emph{polar} 
surfaces or interfaces cannot exist.~\cite{Noguera-08}
Yet, thanks to recent progress in epitaxial growth techniques, nominally 
polar terminations of insulating crystals are now routinely prepared and 
characterized within well-controlled experimental conditions.~\cite{Enterkin-07,Lauritsen-11}
This is possible because, in practice, there are several mechanisms 
available for a polar surface to neutralize the problematic excess charge, 
and possibly become thermodynamically stable.
These include adsorption of foreign gas-phase species, changes in the surface 
stoichiometry, ionic and/or electronic reconstructions, or local metallization
via accumulation of intrinsic free carriers;
each of the above can, 
in principle, prevent the ``polar catastrophe'' by restoring the 
correct charge balance at the surface.

Understanding and controlling these compensation mechanisms is  
a subject of great importance for many areas of fundamental 
science and technology. 
For example, surface polarity is of great interest for catalysis, gas
sensing and energy applications~\cite{Yun-07,Levchenko-08}, as adsorption 
and/or redox of gas-phase species are known to be strongly influenced 
by the electrostatic environment~\cite{Garrity-10}. 
In the context of perovskite-structure thin films and heterostructures,
control of surface charge/polarity is currently investigated as a route 
towards the development of novel field-effect devices (e.g. in the  
LaAlO$_3$/SrTiO$_3$ system~\cite{Cen-08,chargewrite,Son-10}), or 
ferroelectric memories based on the tunneling electroresistance 
effect~\cite{Garcia-09}.
Central to rationalizing all these phenomena is the intimate relationship 
between surface charge and bulk polarization in crystalline insulators, 
which was formally established by Vanderbilt and King-Smith in 
1993.~\cite{Vanderbilt/King-Smith:1994}

From the point of view of the theoretical analysis, it is crucial to 
establish an unambiguous criterion to classify a given surface as 
``polar-compensated'' (i.e. an originally polar surface that was neutralized
via one of the aforementioned mechanisms) or intrinsically non-polar.~\cite{Noguera-08}
This is not just a matter of nomenclature, but has very concrete practical 
relevance: non-polar terminations generally tend to be more stable, as 
extrinsic (i.e. not originated from the primitive building blocks of the  
insulating bulk crystal) sources of compensating charge tend to have a high 
energy cost.
Furthermore, if a given surface is polar, one needs to know precisely how 
much external charge is needed to neutralize it; this greatly facilitates the 
theoretical analysis by restricting the number of possible candidate structures.
We shall see in the following that, while the energetics is a genuine surface 
property, an exact answer to the latter question can be given already at 
the bulk level. Many authors have already addressed this issue in the 
past -- we shall briefly mention hereafter the approaches that are most 
directly relevant to our work.



Tasker~\cite{Tasker} modeled a given ionic crystal as a lattice of point 
charges, corresponding to the nominal valence of the ions.
Based on an abrupt truncation of this lattice, a given surface is then 
classified as polar on non-polar, depending on the behavior of 
the electrostatic energy.
In particular, in the former (polar) case, the bulk repeated unit
cell carries a finite dipole moment; this produces a diverging
electrostatic potential unless compensated by an equal and opposite 
external surface charge density.
This model, despite its simplicity, turned out to be surprisingly effective,
and was able to correctly predict, at least at the qualitative level, the
polar or non-polar nature of the vast majority of insulating surfaces.
However, at the quantitative level this model has clear limitations. 
Many oxides and semiconductors display a marked covalent
character, and the bulk polarization departs significantly from 
the value that can be inferred from atomic positions and nominal 
valence charges. Hence the need for a more accurate treatment.

To address these issues, and adopt a more realistic description of
the charge density of the solid, Goniakowski {\em et al.}~\cite{Noguera-08} 
proposed a different criterion for classifying surfaces as polar or non-polar. 
At the heart of the strategy of Ref.~\onlinecite{Noguera-08} is the concept of 
``dipole-free'' unit cell. Given a certain plane orientation, one
can demonstrate that it is always possible to choose a dipole-free
repeated unit along the normal to that plane; then, the remainder
charge that is left at the surface (once the bulk units have been 
removed) determines the polar or non-polar character of the 
termination.
This criterion, however, is not free from ambiguities: for the same 
surface termination there may exist more than one possible choice of 
the dipole-free bulk unit cell, which might lead to opposite 
conclusions about the polar or non-polar character of the surface
(see section~\ref{sec:dipole-free} for a detailed discussion). 
Also, identifying the dipole-free unit cell might be 
cumbersome in the case of higher index surfaces, where the
structural complexity of the larger cell could complicate
this type of analysis.
Finally, the intuitive appeal of Tasker's model is apparently
lost in the strategy of Ref.~\onlinecite{Noguera-08}: one 
needs to look at the ground-state charge density (e.g.
as provided by a first-principles calculation) before
drawing a conclusion.

There are two further issues that are common to both methods. First,
it is universally agreed that the surface polarity is a property of 
the actual lattice termination. This means that, for a given 
material and surface plane orientation, there might be polar and 
non-polar terminations, depending on the surface stoichiometry.
However, there is no established recipe to unambiguously decide, given 
a compound crystal and a surface orientation, whether a stoichiometric 
$1\times 1$ non-polar termination is allowed at all. Furthermore, it
is not clear how to construct, in general, a non-polar candidate 
structure without relying on a heuristic counting of the layer 
charges.
Second, it was correctly recognized by both Tasker and Goniakowski 
that the issue of surface polarity is directly related to the bulk 
polarization of the material. However, neither model traces a
formal link to the modern theory of polarization in periodic
insulators~\cite{King-Smith/Vanderbilt:1993}, where the 
macroscopic ${\bf P}$ is a multivalued vector field, written in 
terms of the \emph{phases} of the wavefunctions. Only the total 
charge density (modulus of the wavefunctions) is considered in the 
model of Ref.~\onlinecite{Noguera-08}, while explicit electronic 
orbitals are not addressed by Tasker's approach.
Recent theoretical works have indeed highlighted the importance 
of the formal Berry-phase polarization in discussing polarity at
surfaces~\cite{Levchenko-08} and interfaces~\cite{Stengel-09.1}, 
but a general formulation of the problem, based on the formalism 
established in Ref.~\onlinecite{Vanderbilt/King-Smith:1994},
is still missing.

Here we show that a Wannier function representation~\cite{Marzari/Vanderbilt:1997,wannier-05}
together with the ``interface theorem'' of Ref.~\onlinecite{Vanderbilt/King-Smith:1994},
provide a very natural framework for addressing the above issues.
Wannier functions were already shown to be a very useful tool, in
layered superlattices~\cite{Xifan_lp,Xifan_sl}, for partitioning 
the polarization of a crystal into the contribution of individual 
charge-neutral units.
Most importantly, Wannier functions are intimately linked to the 
modern theory of polarization in solids~\cite{ferro:2007}, and therefore 
appear to be the most appropriate ingredient to discuss the issue 
of surface polarity, where the basic question concerns the existence 
of a finite dipole moment perpendicular to the surface plane.
We shall provide, based on this description, precise criteria to 
establish whether truncating a bulk crystal along a given 
crystallographic orientation can yield a non-polar surface.
We shall demonstrate that answering this question involves 
only an analysis of the bulk, and that our scheme naturally
leads to candidate structures that can be used as a starting point
for the subsequent determination of the thermodynamic ground state.
To demonstrate our arguments, we focus on the surfaces of LaAlO$_3$
and SrTiO$_3$, two prototypical perovskite materials that have been 
at the center of the attention in the past few years as their polar
(100) interface exhibits numerous peculiar properties.
%

This work is organized as follows. In section~\ref{sec:theory} we
introduce our definition of polar surface and its formal relationship 
to the theory of bulk polarization. We also establish a direct link to
Tasker's model and we compare it to the ``dipole-free'' cell approach.
In section~\ref{sec:results} we apply this formalism to a variety 
of systems, including non-polar LaAlO$_3$$(n10)$ and SrTiO$_3$(111) 
surfaces. We also discuss electronic/ionic compensation mechanisms of
polar LaAlO$_3$(100).
In section~\ref{sec:discussion} we briefly address some related topics,
including the case of ferroelectric surfaces, and possible extensions to  
covalent semiconductors. 
Finally, in section~\ref{sec:conclusions} we present a brief summary and
the conclusions.

\section{Theory}

\label{sec:theory}

\subsection{Definition of polar surface}

In full generality, for the surface of a crystalline insulator 
to be electrostatically stable, it must have a vanishing density 
of \emph{physical} surface charge, $\sigma_{\rm surf}=0$.
In order to introduce the notion of surface polarity, it is 
useful to separate $\sigma_{\rm surf}$ into two distinct contributions,
and rewrite the stability condition as
\begin{equation}
\sigma_{\rm ext} + {\bf P}_{\rm bulk} \cdot {\hat{n}} = 0.
\label{eq:neutral}
\end{equation}
Here ${\bf P}_{\rm bulk}$ is the bulk polarization, ${\hat{n}}$
is the normal to the surface plane, and $\sigma_{\rm ext}$ is a surface 
density of ``external'' compensating charges, which encompasses
all contributions that cannot conveniently be described as 
``bulk-like'' in nature (we include the latter in ${\bf P}_{\rm bulk}$).
$\sigma_{\rm ext}$ typically includes free charges (e.g. in the form of 
a confined electron gas) and/or bound charges (either in the form of surface 
adsorbates, vacancies, non-stoichiometric reconstructions, or 
non-isoelectronic substitutions).
%
%

We define a given surface as non-polar if the stability criterion 
Eq.~(\ref{eq:neutral}) can be satisfied in the absence of external
charges $\sigma_{\rm ext}$, which implies
\begin{equation}
{\bf P}_{\rm bulk} \cdot {\hat{n}} = 0.
\label{eq:nonpolar}
\end{equation}
This equation, at first sight, looks inconsistent with the current understanding
of the surface polarity problem. It is now widely accepted that the
polar or non-polar attribute is a property of the \emph{termination}, not only of
the material and surface plane orientation, contrary to what Eq.~(\ref{eq:nonpolar}) 
seems to suggest.
We shall see in the following section that the choice of the termination
is only apparently absent from Eq.~(\ref{eq:nonpolar}). It is implicitly 
included through the intrinsically multivalued nature of ${\bf P}_{\rm bulk}$, 
which is a well-established aspect of the modern theory of polarization in
bulk insulators.~\cite{King-Smith/Vanderbilt:1993}

\subsection{The bulk polarization}

\subsubsection{As a Berry phase}

We consider a crystalline insulator described by three primitive 
translation vectors ${\bf a}_{1,...,3}$ and a basis of $N$ atoms
located at positions ${\bf R}_\alpha$, with $\alpha=1,...,N$.
The ``formal''~\cite{ferro:2007} bulk polarization is usually 
defined as
\begin{equation}
{\bf P}_{\rm bulk} = \frac{1}{\Omega} \Big(  \sum_{\alpha=1}^N {\bf R}_\alpha Z_\alpha -
2e \sum_{i=1}^3 \frac{\phi_{\rm el}^{(i)} {\bf a}_i}{2\pi} \Big).
\label{eq:berry}
\end{equation}
Here $Z_\alpha$ is the charge of the ionic core $\alpha$, $e$
is the (positive) electron charge and $\phi_{\rm el}^{(i)}$ is 
the Berry phase~\cite{King-Smith/Vanderbilt:1993} along the reciprocal-space 
vector $i$; for simplicity we assume spin pairing, hence the
factor of $2$ in the electronic contribution.

It is important to note that ${\bf P}_{\rm bulk}$, as defined in 
Eq.~(\ref{eq:berry}) is only defined modulo a ``quantum of polarization'';
in other words, it is not a single-valued but a multi-valued function 
of the electronic and structural degrees of freedom. 
This indeterminacy concerns both the ionic and the electronic parts in
Eq.~(\ref{eq:berry}). On one hand, one has the freedom to choose any
of the periodically repeated images of each atomic specie, and thus 
change ${\bf R}_\alpha$ by an arbitrary translation vector of the
type $\Delta {\bf R} = n_1 {\bf a}_1 + n_2 {\bf a}_2 + n_3 {\bf a}_3$.
On the other hand, $\phi_{\rm el}^{(i)}$ are phases of complex numbers,
and therefore only defined modulo $2 \pi$.

In the following sections we shall use the equivalent formulation of
${\bf P}_{\rm bulk}$ in terms of Wannier functions to illustrate 
the relationship between the multivaluedness of ${\bf P}_{\rm bulk}$ and
the termination of the crystal lattice.

\subsubsection{From the Wannier functions}

We shall explicitly assume, from now on, a single-particle picture,  
in terms of a Kohn-Sham set of orbitals, and a conventional (as opposed
to topological) insulating state. 
Within these assumptions, it is possible to express the electronic 
ground state of the bulk solid in terms of a set of maximally-localized
Wannier functions~\cite{Marzari/Vanderbilt:1997}, which are exponentially 
localized in direct space~\cite{Brouder-07}.
Based on this representation, Eq.~(\ref{eq:berry}) can be rewritten as
\begin{equation}
{\bf P}_{\rm bulk} = \frac{1}{\Omega} \Big(  \sum_{\alpha=1}^N {\bf R}_\alpha Z_\alpha -
2e \sum_{j=1}^{N_{\rm el}/2} \langle {\bf r} \rangle_j \Big),
\label{eq:wannier}
\end{equation}
where $N_{\rm el}$ is the total number of electrons in the primitive cell,
and $\langle {\bf r} \rangle_j$ is the location of the center of the 
$j$-th Wannier function.

Alternatively, we can think in terms of a charge density distribution
that consists in the basis of atomic point charges together with the
Wannier densities,
\begin{equation}
\rho_{\rm cell} ({\bf r}) = \sum_{\alpha=1}^N Z_\alpha \delta({\bf r}-{\bf R}_\alpha)
- 2e \sum_{j=1}^{N_{\rm el}/2} |w_j({\bf r})|^2,
\label{eq:rhocell}
\end{equation}
where $w_j({\bf r})$ is the $j$-th Wannier function of the primitive cell.
By construction, the sum of all the periodic images of $\rho_{\rm cell}$
``tiles'' the total charge density of the extended solid.
Then, by combining Eq.~(\ref{eq:wannier}) and Eq.~(\ref{eq:rhocell}) one 
immediately obtains the intuitive connection to the Clausius-Mossotti formula,
\begin{equation}
{\bf P}_{\rm bulk} = \frac{\bf d}{\Omega},
\end{equation}
where ${\bf d}$ is the dipole moment of $\rho_{\rm cell}$. 

This formulation provides a transparent way to partition the total charge 
density into individual primitive units, whose dipole moment correctly
yields the formal value of ${\bf P}_{\rm bulk}$. 
In doing so, the \emph{phase} indeterminacy of the electronic 
contribution to the polarization discussed in the previous Section has
been reduced to a \emph{lattice} indeterminacy, in all respects 
analogous to that characterizing the ionic contribution. In other words, 
all the complications related to the quantum-mechanical nature of the 
electrons have been mapped into a system of classical point charges, 
where the atoms and the electrons are formally treated on the same footing. 
In the following section we discuss how this Wannier representation
can be further partitioned into smaller units that retain the
chemical information about the formal oxidation state of each ion, 
which is central to the notion of ``polar surface''.

\subsection{Formal ionic charges }

The location of the Wannier functions generally reflects the bonding
properties of the material -- in ionic solids they will cluster around the
atoms, while in covalent materials they will tend to occupy the bond
centers.
We shall assume that the solid has at least a certain degree of ionic
character, so it is possible to ``assign'' each Wannier function to a
given atom without ambiguities; this is certainly true in most known 
oxide materials. (With some caution the ideas developed here can be 
conveniently adapted to any crystalline insulator; we shall briefly 
discuss the example of purely covalent semiconductors in 
Sec.~\ref{sec:discussion}.)
We then combine each ion $\alpha$ with the Wannier orbitals $j$
that ``belong'' to it and define a set of compound charge distributions
that we call ``Wannier ions'' (WI), 
\begin{equation}
\rho_{\rm WI}^{(\alpha)} ({\bf r}) = Z_\alpha \delta({\bf r})
- 2e \sum_{j \in \alpha} |w_j({\bf r + {\bf R}_\alpha})|^2 \Big).
\label{eq:rhowi0}
\end{equation}
(We operated a translation so that the nucleus sits in the origin.)
As the Wannier function locations usually agree remarkably well with 
chemical intuition, each of these $N$ charge distributions will carry 
a monopole $Q_\alpha$ corresponding to the ``nominal'' charge of the 
ion (e.g. $-2e$ for O, $+2e$ for Sr, $+4e$ for Ti).
In addition to their net charge, the WI are non-spherical and 
generally carry non-zero dipole moments ${\bf d}_\alpha$. 
(Higher multipoles are also present, but are not directly 
relevant for the present discussion.) 
$\rho_{\rm cell}({\bf r})$ can now be rewritten in terms of the WI densities,
\begin{equation}
\rho_{\rm cell} ({\bf r}) = \sum_\alpha \rho_{\rm WI}^{(\alpha)} ({\bf r}-{\bf R}_\alpha),
\label{eq:rhowi}
\end{equation}
which is equivalent to Eq.~(\ref{eq:rhocell}) except that here we use
precautions to keep the basic WI units intact.
It follows that the dipole moment of $\rho_{\rm cell}({\bf r})$ can be 
written in terms of two contributions, 
\begin{equation}
{\bf d} = {\bf d}_{\rm PC} + {\bf d}_{\rm WI}.
\label{eqd}
\end{equation}
The first term is the dipole moment of a system of point charges
located at positions ${\bf R}_\alpha$,
\begin{equation}
{\bf d}_{\rm PC}=\sum_\alpha {\bf R}_\alpha Q_\alpha.
\end{equation}
The second term is the sum of the individual dipole moments of
the WI,
\begin{equation}
{\bf d}_{\rm WI}=\sum_\alpha {\bf d}_\alpha,
\end{equation}
It is possible to show that ${\bf d}_{\rm WI}$ is a 
single-valued, gauge-invariant quantity; this number contains all 
the non-trivial electronic contributions to the polarization
that are due to the deformation of the ionic orbitals in the
crystalline environment. 
The gauge invariance of ${\bf d}_{\rm WI}$ might be surprising at 
first sight, as the individual dipole moments ${\bf d}_\alpha$ are 
manifestly gauge-dependent, i.e. they depend on the specific algorithm 
used to localize the Wannier functions.
This arbitrariness cancels out when all ${\bf d}_\alpha$ are summed 
up, as long as the assignment of each Wannier function to a specific
lattice site remains unambiguous.
This is equivalent to stating that a given choice of ${\bf R}_\alpha$
uniquely determines the branch choice of the electronic polarization,
which is a reasonable assumption in ionic materials where the nature
of the valence wavefunction has typically a marked atomic character.

With this new decomposition of ${\bf d}$, we have overcome an important
drawback of Eq.~(\ref{eq:wannier}) and Eq.~(\ref{eq:berry}): in the latter
two equations the decomposition of ${\bf P}_{\rm bulk}$ into ionic and 
electronic contributions is physically meaningless -- only the sum of 
the two terms is well defined. 
Here both ${\bf d}_{\rm PC}$ and ${\bf d}_{\rm WI}$ are formally 
meaningful objects. All the indeterminacy in the definition
of ${\bf d}$ has been recast into the term ${\bf d}_{\rm PC}$, which
has an intuitive interpretation as the dipole moment of a system of 
classical ions, each of them carrying its formal valence charge.
The term which contains the quantum-mechanical information about the
electronic polarization effects is now ${\bf d}_{\rm WI}$, which 
is now a single-valued quantity; this has also a simple physical
interpretation as the total dipole moment of the electronic clouds
of the WI.

Before closing this part, it is useful to point out a direct relationship
between our formalism and the ``layer polarizations'' (LP), $p_l$, introduced in 
Refs.~\onlinecite{Xifan_lp} and~\onlinecite{Xifan_sl}.
If one is interested in layered perovskites with stacking axis along
the (001) direction, it is useful to consider a decomposition of the 
charge density of the ABO$_3$ cell into individual AO and BO$_2$ layers.
In the framework of the present work this implies grouping together the WI
that belong to a given oxide layer.
In particular, the total charge density of each layer $l$ can be then 
written as a sum of all WI densities that belong to layer $l$,
\begin{equation}
\rho_l({\bf r}) = \sum_{\alpha \in l} \rho_{\rm WI}^{(\alpha)} ({\bf r}-{\bf R}_\alpha).
\label{eq:rhol}
\end{equation}
In II-IV perovskites the layers are formally charge-neutral~\cite{Xifan_lp,Xifan_sl}.
$\rho_l({\bf r})$ then carries a well-defined dipole moment, which is directly related to
the LP,
\begin{equation}
p_l = \int \bar{\rho}_l(z) z dz = \frac{1}{S} \sum_{\alpha \in l} 
({\bf d}_\alpha + {\bf R}_\alpha Q_\alpha) \cdot \hat z.
\label{eq:pl}
\end{equation}
(The bar indicates in-plane averaging, and the integral is carried out
along the stacking axis; $S$ is the cell cross-section.)
We shall illustrate this layer-by-layer decomposition 
of the total charge density with practical examples in 
section~\ref{sec:dipole-free}.


\subsection{Crystal termination as a bulk property}

We shall illustrate in this section that the multivaluedness of the
term ${\bf d}_{\rm PC}$ can be formally related to the surface
termination of a semi-infinite crystal. This fact is not new, and
was rigorously established within the modern 
theory of polarization~\cite{Vanderbilt/King-Smith:1994}. Here we discuss the  
implications of the ``interface theorem'' for the electrostatics of
polar surfaces.

Following Goniakowski {\em et al.}~\cite{Noguera-08}, we define a 
\emph{frozen bulk termination} as a surface that is obtained by piling 
up ``bulk unit cells'' without any further electronic or ionic relaxation. 
For the time being, we shall limit our discussion of surface polarity 
to this (somewhat unrealistic) type of surface, that we further specify 
hereafter; we shall make the link to more realistic surface models in 
the next section.
In contrast with Goniakowski {\em et al.}~\cite{Noguera-08}, here we define 
our ``bulk unit cell'' as a charge density distribution that results from a 
superposition of bulk WI, as in Eq.~(\ref{eq:rhowi}).
%
%
Then, we construct the charge density distribution
of the semi-infinite surface system as a superposition 
of $\rho_{\rm cell}({\bf r})$,
\begin{equation}
\rho({\bf r}) = \sum_{{\bf R} \cdot \hat n \leq 0} \rho_{\rm cell}({\bf r-R}),
\label{eq:rhosurf}
\end{equation}
where ${\bf R} = n_1 {\bf a}_1 + n_2 {\bf a}_2 + n_3 {\bf a}_3$ is
a real-space translation vector, and again ${\hat n}$ is the normal
to the surface plane. 
This way of defining a frozen bulk termination has two crucial advantages:
(i) The choice of using the compound WI as our ``elementary particles'' 
naturally ensures that every ion in the surface system will have exactly 
the same formal oxidation state as in the bulk. This is a central point in the
definition of a polar surface -- if we allowed for fractional orbital occupations
no surface would be polar. (ii) The choice of cleaving
the \emph{Bravais} lattice, rather than the \emph{crystal} lattice is
particularly advantageous, as it naturally preserves the bulk stoichiometry
everywhere in the system (if we allowed for stoichiometry changes of reconstructions
again the notion of polar surface would be inconsistent). 

It can be easily verified that several types of unreconstructed surfaces can be 
generated by using Eq.~(\ref{eq:rhosurf}), simply by changing the definition
of $\rho_{\rm cell}({\bf r})$. In particular, we have the freedom to construct
$\rho_{\rm cell}({\bf r})$ in many different ways, simply by shifting each WI 
in the basis by an arbitrary Bravais lattice vector $\Delta {\bf R}$.
Thus, the choice of the basis vectors ${\bf r}_i$ uniquely determines the
surface structure, according to Eq.~(\ref{eq:rhowi}) and Eq.~(\ref{eq:rhosurf}).
(Of course, different choices of ${\bf r}_i$ can lead to the same termination;
the ``$\{{\bf r}_i\} \rightarrow $ termination'' relationship is a many-to-one
function.)
On the other hand, we have shown in the previous section that the choice of
${\bf r}_i$ uniquely determines the value of ${\bf P}_{\rm bulk}$, out of
the infinite possibilities allowed precisely by the arbitrariness in the
choice of the basis vectors.
This formally establishes the relationship between ${\bf P}_{\rm bulk}$ 
and the termination of the lattice.
By construction, the \emph{physical} net charge that lies at the surface of 
a frozen bulk termination as defined above is simply  
$\sigma_{\rm surf} = {\bf P}_{\rm bulk} \cdot \hat n$, where 
${\bf P}_{\rm bulk}$ is the dipole moment (per unit volume) of an 
appropriate bulk unit cell, i.e. one that tiles the semi-infinite
solid according to Eq.~(\ref{eq:rhosurf}).

%
%

Within these assumptions, we define a frozen bulk termination polar if the 
bulk building block used to construct it has a net dipole perpendicular to 
the surface plane; we define it non-polar otherwise.
The problem of determining whether a surface is polar or not
is, therefore, reduced to the problem of calculating the dipole moment of a 
bulk unit cell made of WI. This, in turn, can be directly related to the
result of a Berry-phase calculation in the bulk crystal, which can be 
routinely performed with most publicly available codes. In other words, 
the termination itself can be understood as a \emph{bulk} property.

\subsection{Frozen and relaxed surfaces}

It might appear artificial to consider surfaces
that are constructed by stacking electronic orbitals corresponding 
to bulk Wannier functions. 
At a real surface, electronic states always depart from their 
bulk counterparts because of the peculiar chemical and electrostatic
environment produced by the truncation of the crystal.
Furthermore, also the ionic lattice undergoes nontrivial structural 
relaxations in the surface layers, in response to the perturbation of the 
bonding network.
A central point of our formalism is that both (electronic and ionic) 
surface relaxation effects are essentially irrelevant in the context of 
deciding whether a given surface is polar on non-polar. 
As a matter of fact, either type of relaxation only affects the surface
dipole moment, and not the surface charge density.
Thus, genuine surface properties (e.g. the alignment between the bulk bands 
and the vacuum levels, or the surface energy) 
certainly depend on these mechanisms, but the polarity (which 
depends only on the physical surface charge) won't be affected.
This formally establishes the surface polarity as a property that 
can be completely understood at the bulk level -- note that the
termination dependence can also be understood as a bulk property
as specified in the previous section.
Then, all mechanisms that alter the surface \emph{charge} (either in the
form of a local composition change or as a modification of the 
formal oxidation state of the surface ions) are unambiguously 
understood as external compensation effects, and enter the definition
of $\sigma_{\rm ext}$.

\subsection{Construction of arbitrary non-polar terminations}
\label{sec:questions}

So far we have addressed the question of deciding whether a 
given surface, of a certain orientation and termination, is polar
or non-polar.
One could wonder now, for a given bulk compound, 
(i) whether non-polar terminations can be constructed at all; (ii) if yes
along which surface plane orientation;
finally, it would be helpful to (iii) identify candidate non-polar surface
structures based only on bulk information.
In this section we shall illustrate how this is done within the
present definition of surface polarity.

Essentially, the question (i) boils down to finding all possible values 
of ${\bf P}_{\rm bulk}$. This is, within the modern theory of polarization,
a periodic lattice of points. The difference between two arbitrary values 
of ${\bf P}$ is a multiple of a real-space primitive translation vector,
\begin{equation}
{\bf P}'_{\rm bulk} - {\bf P}_{\rm bulk} = \frac{Q_0}{\Omega} (i {\bf a}_1 +
j {\bf a}_2 + k {\bf a}_3).
\end{equation}
Here $Q_0=ne$ is an integer $n$ times the electron charge $e$. $n$, which 
determines the resolution of the ${\bf P}_{\rm bulk}$ mesh depends on the 
convention of how the Wannier functions and the ion cores are grouped together.
In particular, the constraint adopted here of assigning each Wannier 
function to a specific ionic site generally restricts the lattice of 
possible values of ${\bf P}_{\rm bulk}$ to a subset of those allowed by 
Eq.~(\ref{eq:wannier}) and Eq.~(\ref{eq:berry}).
This can be understood by observing that the new elementary building
blocks of the lattice are the ``compound objects'' WI, rather than single 
electrons or ions. 

Answering question (ii) consists in finding the intersections between the 
infinite lattice of ${\bf P}_{\rm bulk}$ values and a given surface 
plane, which is a straightforward geometrical problem.

Answering question (iii) then is easy by recalling the direct relationship
between a given value of ${\bf P}_{\rm bulk}$ and the dipole moment
of a well defined bulk unit.
More specifically, once a value (or a subset of values) of ${\bf P}_{\rm bulk}$ is
found for which ${\bf P}_{\rm bulk} \cdot \hat n = 0$, models of the
non-polar surface can be readily built by stacking [using Eq.~(\ref{eq:rhosurf})]
bulk unit cells that correspond to those same values of ${\bf P}_{\rm bulk}$.
We shall present several practical examples of this strategy in Sec.~\ref{sec:results}.

\subsection{Relationship to previous approaches}

\subsubsection{Tasker model}

\begin{figure*}
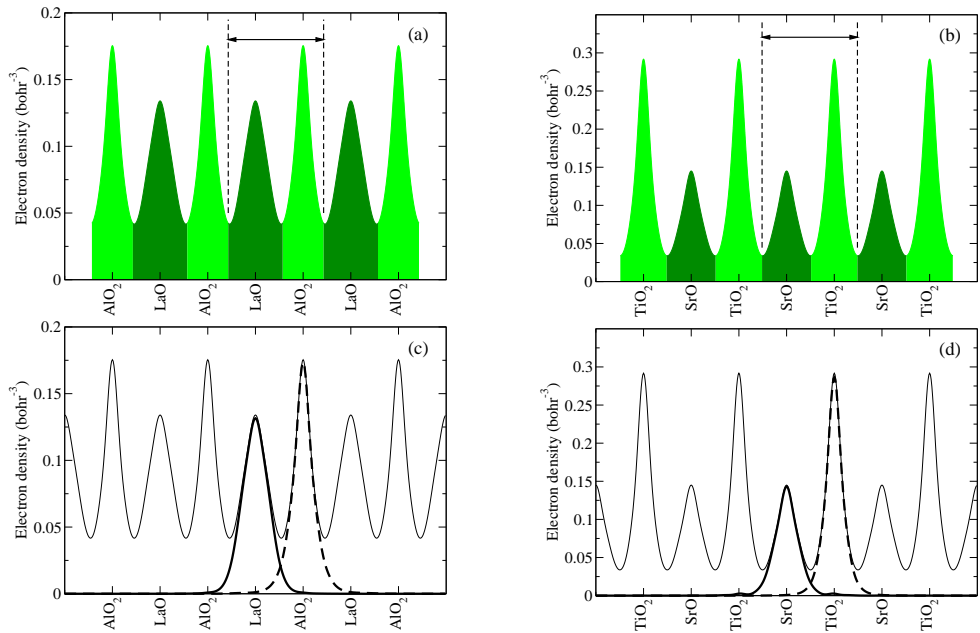

\begin{center}
\includegraphics[width=2.3in,clip]{fig1a.eps} \hspace{1cm}
\includegraphics[width=2.3in,clip]{fig1b.eps} \\
\includegraphics[width=2.3in,clip]{fig1c.eps} \hspace{1cm}
\includegraphics[width=2.3in,clip]{fig1d.eps}
\end{center}

\caption{(a-b) Decomposition of the total valence charge density 
according to the ``dipole-free unit cell'' picture. Light and 
dark shadings indicate the portions that belong to the BO$_2$ 
and AO layers, respectively. We impose both distributions
to be symmetric around the respective atomic layer locations,
and to contain a number of electrons equal to the total valence
charge of the ions. The unit cell indicated by the arrow and the
dashed lines has zero dipole moment by construction in both
LaAlO$_3$ (a) and SrTiO$_3$ (b). (c-d) Decomposition
of the total valence charge based on maximally localized Wannier
functions. The total Wannier densities of the AO and BO$_2$ layers
are shown by thick solid and thick dashed curves, respectively. The
total charge density (obtained by the superposition of the periodically 
repeated Wannier densities) is shown as a thin solid line. \label{fig1}}
\end{figure*}

Eq.~(\ref{eqd}) constitutes the rigorous link between Tasker's model~\cite{Tasker}
and the modern theory of polarization in periodic insulators. 
Within our formalism, the total excess charge at a frozen bulk
termination can be exactly written as
\begin{equation}
\sigma_{\rm surf} = {\bf P}_{\rm bulk} \cdot \hat n = \frac{ (\sum_\alpha {\bf R}_\alpha Q_\alpha + 
{\bf d}_{\rm WI}) \cdot \hat{n}}{\Omega},
\label{eqtasker}
\end{equation}
where the sum is extended over all atoms in the semi-infinite crystal.
%
%
%
The only difference between Tasker's model and Eq.~\ref{eqtasker} is the 
additional, purely electronic contribution ${\bf d}_{\rm WI}$, 
which comes from the polarization of the Wannier ions in the 
crystalline environment.
This contribution vanishes in all solids that are characterized by a center
of symmetry; in these materials the discussion of the surface polarity 
problem in terms of nominal charges is therefore rigorous and exact.
Even in materials where ${\bf d}_{\rm WI} \neq 0$, neglecting 
this term is usually not crucial to assessing the polar or 
non-polar nature of a given surface.
However, considering the WI contribution is essential for a quantitative
estimation of $\sigma$ (which is the excess charge that needs to be 
compensated); this is especially true in ferroelectric materials, which 
generally have a large anomalous contribution to ${\bf P}$.
Thus, our formalism provides a formal justification to Tasker's model, 
and completes it by introducing an additional well-defined electronic 
dipolar contribution, ${\bf d}_{\rm WI}$. 

In addition to this, our strategy has important practical advantages. 
Tasker's approach involves a direct calculation of Eq.(\ref{eqtasker}) 
by means of an infinite lattice sum, whose convergence is ensured by using
Ewald summation techniques. 
This procedure might be cumbersome in practice, and it requires a 
specialized computer code to perform the calculation.
Our strategy greatly simplifies the problem, by reducing it to the
calculation of the dipole moment of a small set of point charges.
This can be done with paper and pencil in few minutes for a surface
of arbitrary orientation, provided that one knows ${\bf d}_{\rm WI}$.
This vanishes in many cases of practical interest -- whenever it
doesn't vanish, only a single bulk Berry-phase calculation is needed to 
evaluate this contribution.
%
Moreover, our strategy allows one to easily answer a number of
physical questions that were difficult to address within Tasker's 
approach, e.g. those discussed in the previous section.
%
%



\subsubsection{Other approaches}

\label{sec:dipole-free}
%
%
In order to fully appreciate the advantages of our
formalism, it is useful to compare it, in a practical case, with the 
alternative notion of ``dipole-free unit cell'' proposed by Goniakowski 
{\em et al.}~\cite{Noguera-08}
For illustrative purposes, we consider the (100) surfaces of two 
prototypical perovskite materials, LaAlO$_3$ and SrTiO$_3$, in 
their cubic high-symmetry phase.
We shall use either the arguments developed in the previous 
sections, based solely on bulk properties of the respective materials,
or the theory of Ref.~\onlinecite{Noguera-08}, in order to assess 
the polar or non-polar character of these surfaces.

In Fig.~\ref{fig1} we plot, along the (001)-oriented $z$ axis, 
the $xy$-planar average of the total valence charge density. 
(The left panels refer to LaAlO$_3$, the right ones to SrTiO$_3$.)
The upper panels show a possible decomposition of the electronic 
charge that leads to a dipole-free unit cell, which we construct
as follows.
First, we count the total charge of the ionic cores of the individual
oxide layers. With the pseudopotentials used in this work, these are
LaO(+17), AlO$_2$(+15), SrO(+16) and TiO$_2$(+24). 
Next, we decompose the total valence charge by cutting it with abrupt
(001) planes located in the interstitial regions. The location of those 
planes is chosen as to (i) respect the inversion symmetry of the crystal,
and to (ii) assign to each layer an electron density that exactly cancels
the positive core charge of that layer.
The resulting electron charge assigned to the AO layers is highlighted 
with a dark shading (light for the BO$_2$ layers).
By construction, the ``unit cell'' obtained by combining two adjacent 
layers (evidenced by the arrow and dashed lines in the figure)
has zero dipole moment in both LaAlO$_3$ and SrTiO$_3$. Hence, this 
construction fails at detecting any fundamental difference between 
LaAlO$_3$ and SrTiO$_3$: both are predicted to have non-polar (001) surfaces.
Of course, this prediction relies on a completely arbitrary partition
of the total electronic charge density. There are many other ways to
do it. For example, if one chooses a different location of the cut
planes (e.g. at the mid-point distance between the atomic planes), 
or yet a more sophisticated prescription (e.g. based on the Bader 
analysis), one generally gets a non-vanishing layer charge in both 
LaAlO$_3$ and SrTiO$_3$. From this perspective, one would have to
conclude that the (001) surfaces of both materials are polar.
The main point that we want to stress here is that, if we base our
analysis solely on the total electronic density [as we have done
in Fig.~\ref{fig1}(a-b)], (i) the choice between one partitioning scheme 
and the other is arbitrary; (ii) any statement about the surface polarity 
inferred from such a partitioning is ambiguous; and (iii) such an analysis
cannot be linked in any ways to the bulk polarization of the material
(the latter cannot be defined, even in principle, in terms of the total 
charge density of a periodic crystal).

In Fig.~\ref{fig1}(c-d) we demonstrate how the Wannier-based decomposition
of the valence density solves this problem. The thin solid lines show, as
above, the ground-state electronic charge densities $\bar{\rho}(z)$ (again,
the bar symbol on $\rho$ indicates that an in-plane averaging was 
performed). The total Wannier densities of each layer, defined as 
$\bar{\rho}_l$ in Eq.~(\ref{eq:rhol}), are shown as thick lines 
(solid for the AO layers, dashed for the BO$_2$ layers). Note that
we show the electron density as positive, and we omit the bare 
pseudopotential charges from the plots. (These are a lattice of Dirac 
delta functions, centered at the oxide layer locations.) As the Wannier 
functions are discrete objects, the total electronic charges are integer 
numbers. Most importantly, the Wannier functions carry some crucial information 
(that is absent in the total valence density) on how the localized
bound charges are organized in the insulating state of each compound. 
It turns out that (summing up the contributions from the cores) the LaO and 
AlO$_2$ layers have a total charge of +1 and -1, respectively, while
the SrO and TiO$_2$ layers result charge-neutral, in perfect agreement 
with the naive assumption of perfect ionicity.
Thus, the Wannier decomposition correctly identifies LaAlO$_3$(001) 
as polar and SrTiO$_3$ as non-polar, in agreement with Tasker's 
classification.

To corroborate our arguments, a further consideration is in order.
One could be tempted to criticize our reasoning by observing that
Wannier functions are by no means uniquely defined starting from a
given set of Bloch orbitals. In Fig.~\ref{fig1}(c-d) we have
chosen a maximally localized~\cite{Marzari/Vanderbilt:1997} representation,
but there is nothing really fundamental behind this choice. Is
the identification of a surface as polar or non-polar robust
against this arbitrariness? The answer is yes. The formal
proof of this statement was derived in 1993~\cite{Vanderbilt/King-Smith:1994},
several years before the maximally localized Wannier functions were 
first introduced. Our choice of a maximally localized representation
is motivated by its intuitive relationship to elementary chemical concepts
(e.g. formal valence), but the reader should keep in mind that this is
just a convenient way of expressing a concept that has solid mathematical 
grounds. In particular, the exact value of the surface charge can be
computed at the bulk level, regardless of the degree of ionicity of the
material.~\cite{Vanderbilt/King-Smith:1994}


\section{Application to perovskite surfaces}

\label{sec:results}

\subsection{LaAlO$_3$ and SrTiO$_3$: bulk properties}

We now use two prototypical perovskite materials, LaAlO$_3$ and SrTiO$_3$,
to illustrate our strategy in practice. This choice of  materials
is motivated by the recent discovery of a conducting electron gas at 
their polar (100) interface.~\cite{Ohtomo-04} This has generated a lively 
excitement in the research community and a renewed interest in the 
theoretical foundations of the surface/interface polarity problem.~\cite{Hwang-book,Bristowe-11,
Stengel-09.1}

We shall address questions (i-iii) raised in Sec.~\ref{sec:questions}.
For simplicity, in both materials we consider only surfaces of the type 
(0ij). This means that only the projection of ${\bf P}_{\rm bulk}$ on the
$yz$ plane is relevant, and our procedure can be conveniently represented
on 2D graphs. Our strategy is general, and this choice was made only 
to simplify the notation and the graphical representations.
We also consider both bulk compounds within their high symmetry
cubic phase. (Both materials are characterized by zone-boundary distortions,
related to rotations and tilts of the oxygen octahedral network; however,
as these distortions are non-polar in nature, they are irrelevant for the 
present discussion.)

\begin{figure}
\includegraphics[width=3.0in]{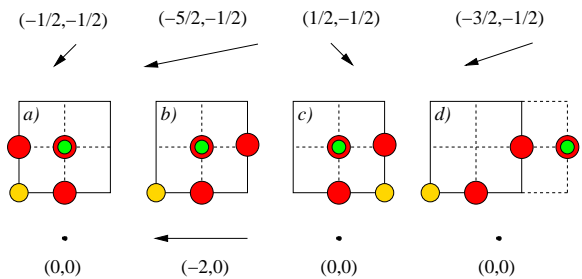}
\caption{Relationship between ${\bf P}_{\rm bulk}$ and the dipole
moment of the bulk unit cell. Red balls represent O ions, with charge 
$Q_{\rm O} = -2e$. Gold balls represent the A-site cation, either 
Sr ($Q_{\rm Sr} = +2e$) or La ($Q_{\rm La} = +2e$); green balls are
the B-site cation, either Ti ($Q_{\rm Ti} = +4e$) or Al ($Q_{\rm Al} = +3e$).
The sketches (a-d) represent the projection of the atomic positions onto a 
(100)-oriented plane. On the top and the bottom are shown the values
of ${\bf P}_{\rm bulk}$ (in units of $e/a_0^2$, where $a_0$ is the 
lattice parameter in either material) resulting in LaAlO$_3$ and
SrTiO$_3$, respectively, from each cell arrangement. \label{fig:dipoles}}
\end{figure}

\begin{figure}
\includegraphics[width=3.0in]{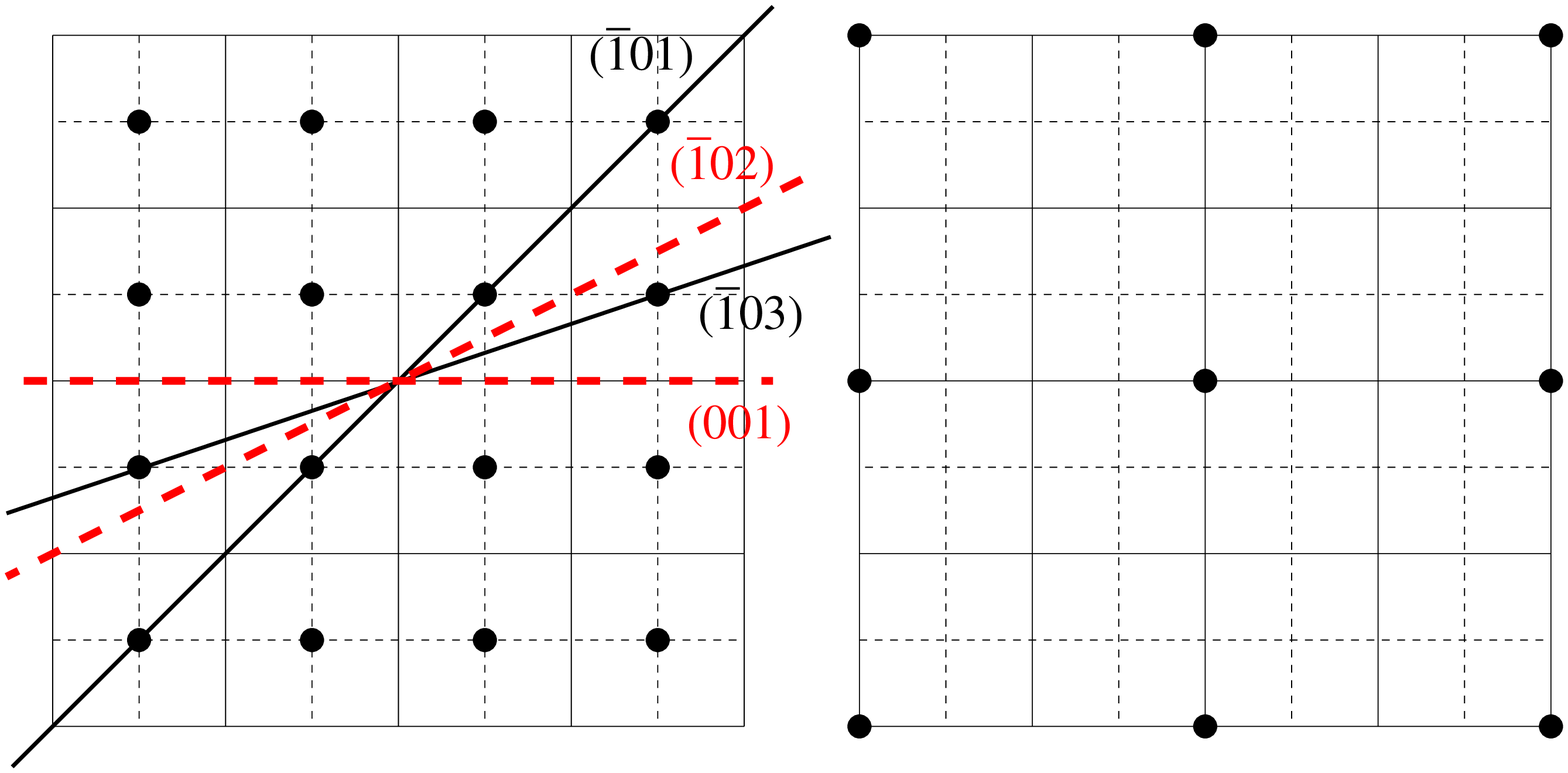}
\caption{Lattices of allowed values of ${\bf P}_{\rm bulk}$ in 
LaAlO$_3$ (left) and SrTiO$_3$ (right). In the left panel we 
show four possible surface plane orientations. The polar orientations
(dashed red lines) have no intersections with the ${\bf P}_{\rm bulk}$ 
lattice. The contrary is true for the non-polar orientation (solid
black lines). \label{fig:platt}}
\end{figure}

To start with (question i), we need to find the lattice of ``allowed'' values of
${\bf P}_{\rm bulk}$ in either material. Recalling that both compounds
are characterized by a center of symmetry, it follows that ${\bf d}_{\rm WI}=0$,
and ${\bf P}_{\rm bulk}$ is exactly determined by the formal valence charges
of the participating ions, all sitting in their high-symmetry lattice sites.
Now, the formal ionic charges are La(+3), Al(+3) in LaAlO$_3$, Sr(+2), Ti(+4)
in SrTiO$_3$; oxygens in either compound have a formal charge of (-2). 
In Fig.~\ref{fig:dipoles} we show how different choices of the crystal basis
of five atoms lead to different dipole moments per unit cell, and hence 
to a different formal polarization.
If we could take all (infinite) combinations, we would obtain an
infinite lattice of points, which is isomorphic with the 
real-space Bravais lattice of the cubic crystal.
The 2D projection of the lattice of ${\bf P}_{\rm bulk}$ in either 
compound is shown in Fig.~\ref{fig:platt}. 
Even if the two compounds are isostructural (recall that we consider both
LAO and STO in their cubic phase), there are two important differences in 
their formal polarization lattice.
First, ${\bf P}_{\rm STO}$ is centered in the origin, while 
${\bf P}_{\rm LAO}$ is centered in $(1/2,1/2,1/2)$. Second, 
${\bf P}_{\rm STO}$ has a coarser mesh than ${\bf P}_{\rm LAO}$
 -- the spacings are doubled because the constituent point charges 
are all even in the former.
Note that both ${\bf P}_{\rm bulk}$ lattices are centrosymmetric,
consistent with the absence of a spontaneous polarization in 
either material.~\cite{ferro:2007}

To answer question (ii), we need to find all possible intersections
between a surface plane and the allowed values of ${\bf P}_{\rm bulk}$;
the projection of a few representative surface planes are plotted
in Fig.~\ref{fig:platt}(a).
As it can be readily appreciated from the diagram, the aforementioned 
qualitative differences between the respective ${\bf P}_{\rm bulk}$ 
lattices of STO and LAO have important consequences on the electrostatics
of the surfaces.
In particular, in STO the origin belongs to the allowed values of 
${\bf P}_{\rm STO}$, and any surface plane intersects the origin
by construction; therefore, a non-polar surface of any possible 
orientation can be readily constructed [we shall illustrate the case 
of STO(111) in Sec.~\ref{sec:sto111}].
Conversely, in LAO only specific plane orientations intersect the 
${\bf P}_{\rm LAO}$ lattice [note that the (100) orientation
is correctly classified as polar]. In the following Section we shall 
consider a subset of these (infinite) possibilities, i.e. the 
vicinal $(01n)$ surfaces, where $n$ is an arbitrary odd integer number.
We shall focus on the lowest-index cases with $n=1,3,5$.

\subsection{Vicinal LaAlO$_3$ surfaces}
\label{sec:vicinal}

\begin{figure}
\begin{center}
\includegraphics[width=2.0in]{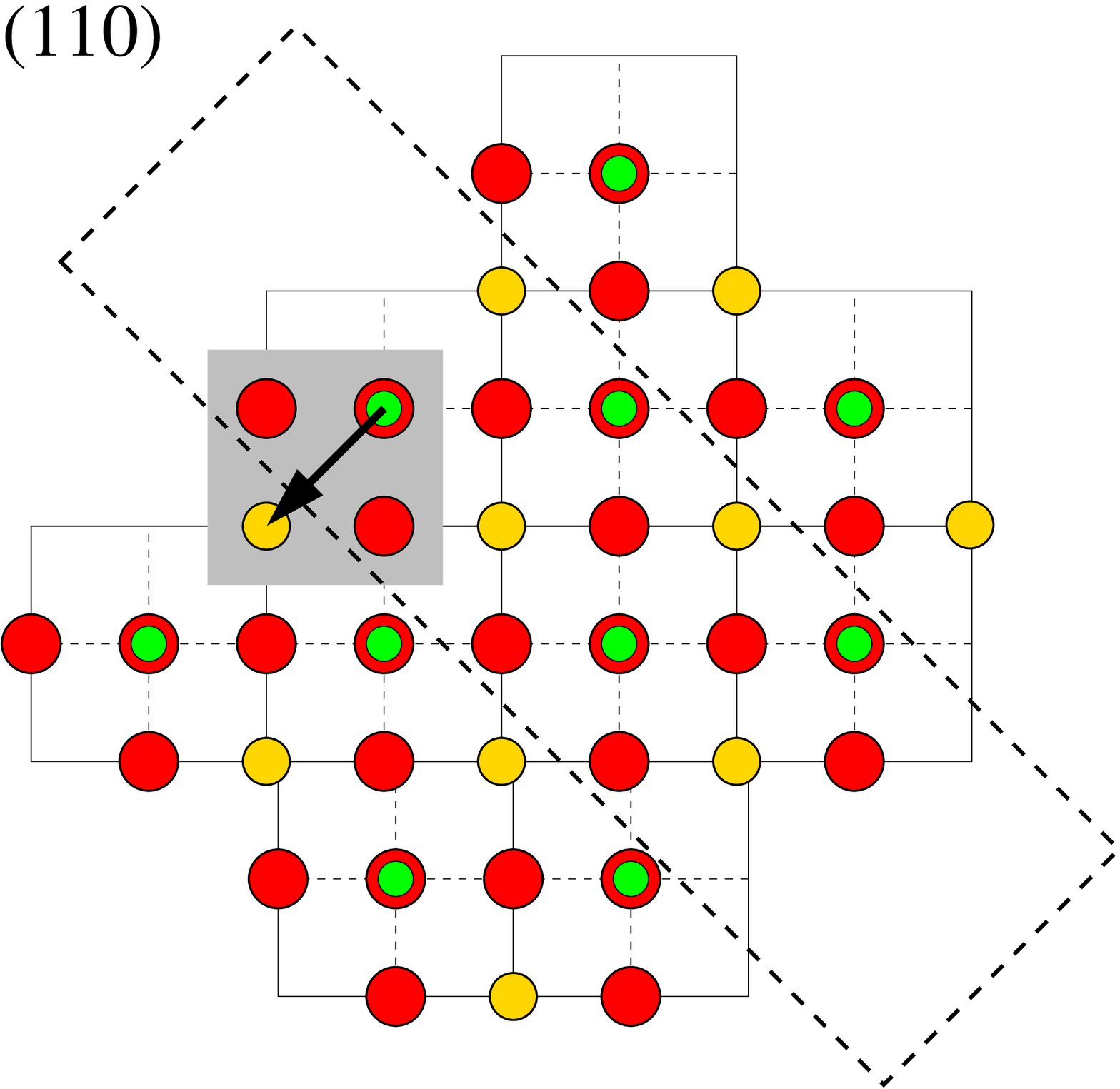} \\
\vspace{10pt}
\includegraphics[width=2.2in]{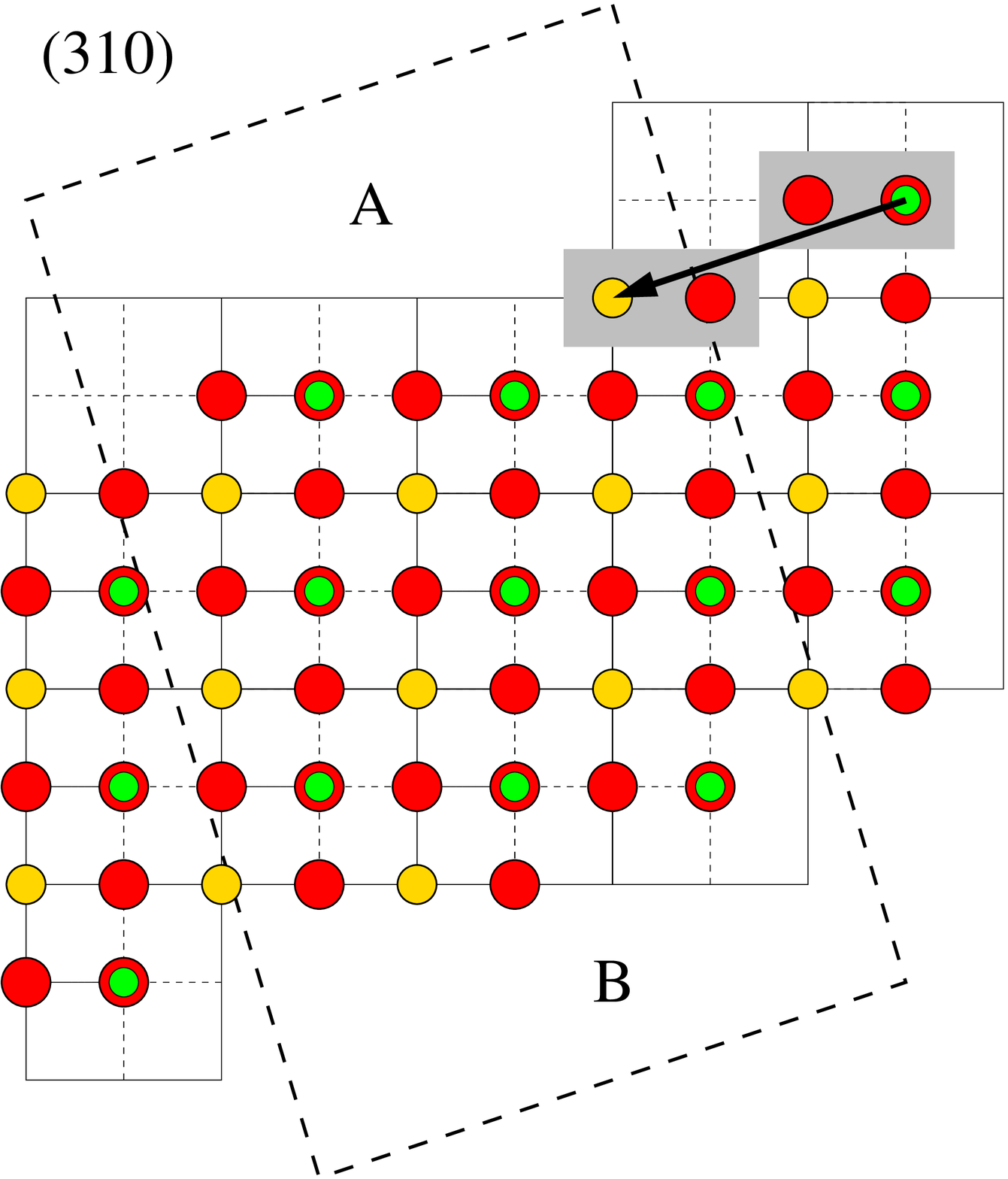} \\
\includegraphics[width=2.8in]{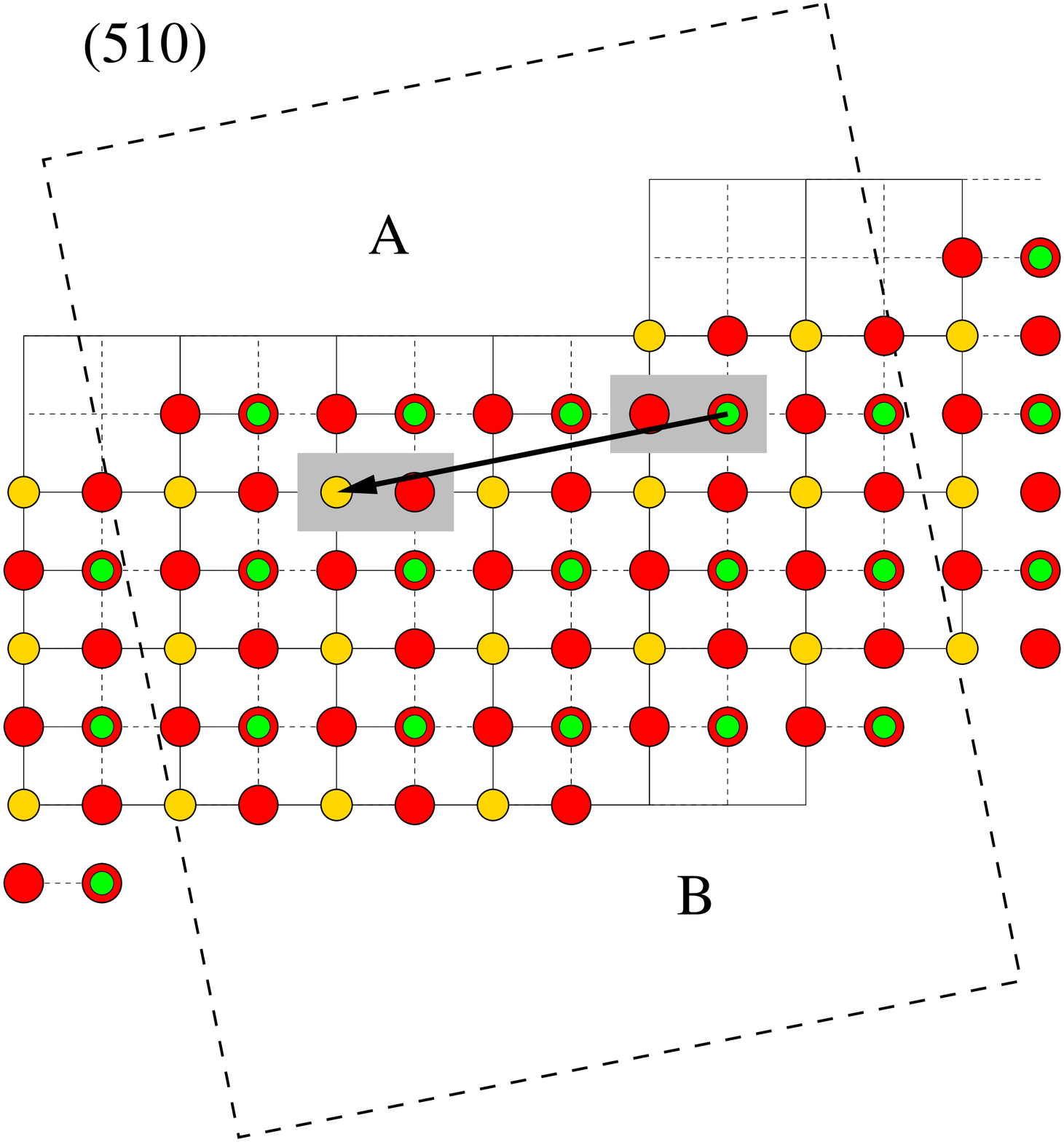} 
\caption{Slab models for the vicinal LAO surfaces described
in the text. Top: (011). Center: (013). Bottom: (015). Color code
for atoms is the same as in Fig.~\ref{fig:dipoles}. Thick dashed
lines indicate the supercells used in the simulations. Thin 
lines are guides to the eye. Shaded areas highlight the basic 
primitive unit that was used to construct the slab model. Thick
arrows indicate the dipole moment of the basic unit, parallel 
to the surface plane. \label{fig:models}}
\end{center}
\end{figure}

We first construct preliminary models for the $(01n)$ surfaces
with $n=1,3,5$. These are obtained by using slab geometries,
within the supercell method. First we choose a unit cell with
the appropriate translational periodicity in plane, and enough 
room along the out-of-plane direction to accommodate both the
slab and a vacuum region (slab and vacuum thicknesses are treated
as convergence parameters). 
Second, we tile the slab region with repeated copies of 
a well-defined primitive basis of atoms, which is chosen in
a such a way that its dipole moment lies
exactly parallel to the surface plane. (This implies that 
the choice of the basis depends on the surface orientation.)
This procedure leads to the slab models sketched in Fig.~\ref{fig:models}.
%

\begin{table}
\begin{ruledtabular}
\begin{tabular}{c|ccc}
Surface type & \multicolumn{3}{c}{Surface orientation} \\
 &    (011) & (013) & (015) \\
\hline
A &        1.93  &  2.23 & 2.15 \\
B &        1.93  &  1.85 & 1.83 \\
\hline
Cleavage & 3.86  &  4.08 & 3.98
\end{tabular}
\end{ruledtabular}
\caption{Calculated energy per area for the LAO surfaces described in
 the text. An ideal cleavage of the crystal is assumed to leave a pair 
of A and B surfaces. All values are in J/m$^2$. \label{tab1} }
\end{table}

The first observation is that all these surface models
[except maybe the (011) case] present alternating LaO-type
and AlO$_2$-type terraces, and these terraces tend to grow
wider and wider for increasing $n$. 
Note that [again, with the only exception of the (011) case],
the construction described above produces, in fact, \emph{two} 
inequivalent surface structures for each orientation. In other 
words, the models of Fig.~\ref{fig:models} do not enjoy inversion 
symmetry. 
We shall refer to these two surfaces as ``type A'' and ``type B'',
where type A presents LaO-type step edges and type B has AlO-type 
edges.
Remarkably, it is easy to realize that one can change from A-type
to B-type simply by displacing an oxygen atom from one step edge
to the neighboring one. This way, starting from the ``mixed''
AB-type slabs Fig.~\ref{fig:models} one can readily construct
pure AA or BB slabs. One can verify that the resulting AA and BB models 
do enjoy inversion symmetry.
Since going from A to B preserves the bulk stoichiometry, this
allows for a rigorous definition of the surface energy for all 
individual surface structures.

In practice, in the simulations we use a slab thickness of 
approximately 4-5 LAO cells in each case, which is more than 
sufficient to obtain a well-converged value of the surface 
energy.
The surface energy is defined as
\begin{equation}
E_{\rm surf} = \frac{1}{2S} (E_{\rm slab} - N E_{\rm bulk}),
\end{equation}
where $E_{\rm slab}$ and $E_{\rm bulk}$ are the relaxed total energy of
the slab supercell and of LAO bulk, $N$ is the total number of LAO units in
the slab model, and $S=a_0^2\sqrt{1+n^2} $ is the surface area in each case
($a_0$ is the lattice parameter of cubic LAO; the factor of 2 takes into
account the fact that a slab has two surfaces).
In Table~\ref{tab1} we report the results. Comparing these values with 
previous literature studies is difficult, as studies of vicinal perovskite
surfaces are scarce. Only the lowest-index (011) surface type has been 
investigated to some extent, although we weren't able to find data
specific to LAO. Concerning other perovskite materials, Eglitis and 
Vanderbilt~\cite{Eglitis:2008} reported an energy of 1.52 J/m$^2$ for 
an isostructural O-terminated SrTiO$_3$(011) surface structure. The value 
we obtain for LAO, 1.93 J/m$^2$, is somewhat larger but otherwise of 
comparable magnitude. Note that in the study of Ref.~\onlinecite{Eglitis:2008} 
a different (hybrid) functional was used -- LDA might well overestimate 
surface energy values due to the well-known overbinding issues.

It is interesting to note that in the case of B-type surfaces the energy
decreases slightly for increasing index $n$. We ascribe this behavior 
to the lower steps-to-terraces ratio in the (013) and (015) surfaces
(undercoordinated step sites are likely to be less favorable). We
consider it unlikely, however, that this energy be further
reduced for $n>5$. Increasing $n$ would lead to larger and larger terraces
that are locally charged [either of the LaO(+) or AlO$_2$(-) type],
and the electrostatic cost (roughly linear in $n$) would eventually dominate 
over the step energy (proportional to $1/n$) in a way that bears many 
analogies to Kittel's theory of domain walls.
Still, the increased stability of the vicinal (013) and (015) surfaces
[compared to the (011) orientation] suggests that these geometries could
be, in principle, fabricated under appropriate experimental conditions.
The simultaneous presence AO and BO$_2$ domains appears promising for 
applications, e.g. in selective self-assembly of functional nanostructures,
as it was recently shown in the case of SrTiO$_3$~\cite{Bachelet:2009}

\begin{figure}
\begin{center}
\includegraphics[width=3.2in,clip]{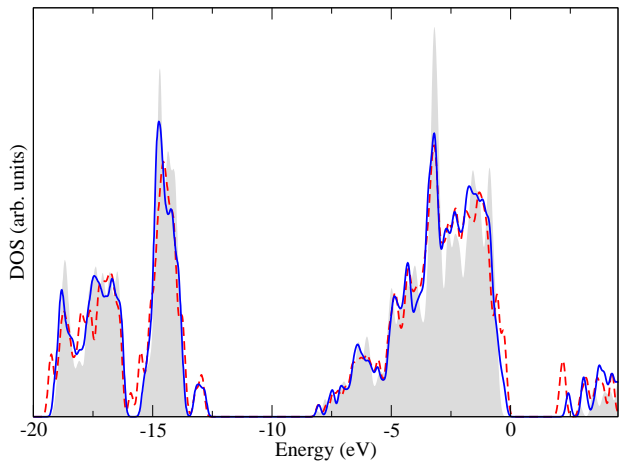}
\caption{Total density of states of the LAO(013) slab models. 
Surface A (black curve), B (red dashed) and bulk (shaded 
area) are shown. \label{fig:ldos} }
\end{center}
\end{figure}

The above considerations on the energetics haven't answered an
important question yet: how can we verify that these  
surfaces are indeed non-polar, consistent with our predictions? A
useful indication comes from the density of states. If a surface
is polar, then there is a need for compensation via additional charge
carriers (either electron or holes) that deplete or populate the
energy bands of the crystal in proximity of the problematic termination. 
This typically results in a metallic surface. Conversely, if the surface 
is non-polar, the bulk-derived Wannier functions alone are sufficient to 
ensure electrostatic stability, and therefore the system can remain 
\emph{insulating}.
In Fig.~\ref{fig:ldos} we show the total density of states extracted 
from a (013) (A- or B-type) slab model, compared with the bulk LAO
density of states. 
In all cases there is a wide gap separating the unoccupied from the 
occupied states. This fact, together with the inversion symmetry and 
perfect bulk stoichiometry of the slabs, directly demonstrates that the
surfaces are non-polar, and that every atom contributes with a total
number of electrons that exactly corresponds to its formal ionic
valence. Similar considerations apply to the (110) and (510) surface
models (not shown).

It is important to stress that, contrary to a common misconception,
all these surfaces are perfectly \emph{stoichiometric} by 
construction, and they are \emph{non-reconstructed} as they
have the highest possible translational symmetry that is allowed
by each plane orientation. 
It is often assumed that the only ``legitimate'' structures
that can be named frozen bulk terminations are those that are 
obtained upon cleavage of the \emph{crystal} lattice, i.e. 
preserving the integrity of the bulk-like atomic planes.
This is, however, just a convention that has nothing fundamental to 
it. We believe it is more practical to truncate the \emph{Bravais}
lattice instead. This automatically preserves stoichiometry and
translational symmetry, and dramatically simplifies the description
of surface electrostatics.
%

As a final remark, it is fairly easy to realize that all the $(01n)$
surface models presented in this section are non-polar for \emph{any} 
non-ferroelectric perovskite material (or for a ferroelectric one
in its high-temperature symmetric phase). 
This can be simply understood by observing that, by replacing the
cations in each bulk primitive basis (see Fig.~\ref{fig:models}) 
with those of a different charge family (i.e. I-V or II-IV), the 
dipole moment changes its magnitude but not its direction.  Also, all 
the surface models enjoy inversion symmetry and have ideal bulk 
stoichiometry -- the absence of a dipole moment normal to the 
surface plane is therefore automatically guaranteed.
Therefore, many of the considerations made here in relationship to 
the specific LAO case are actually completely general, and apply to
all cubic perovskite compounds.


\subsection{Non-polar SrTiO$_3$(111) surfaces}
\label{sec:sto111}

To further illustrate our arguments, we move now to the case of
SrTiO$_3$. According to our definition of polar surface, and by 
observing that the simplest choice of the SrTiO$_3$ bulk unit has
zero dipole moment, one would conclude that in SrTiO$_3$ \emph{any} 
surface orientation is ``non-polar''. To illustrate this point we shall 
consider here the (111) surface, which has been classified as
polar by most authors. Note that this surface is indeed polar if one 
insists on terminating the crystal lattice with either a Ti or a 
SrO$_3$ layer. Our prescription of cleaving the Bravais lattice and
tiling it with well-defined bulk-like formula units is less restrictive,
and allows for non-polar terminations as we shall see in the following.

\begin{figure}
\begin{center}
\includegraphics[width=1in,clip]{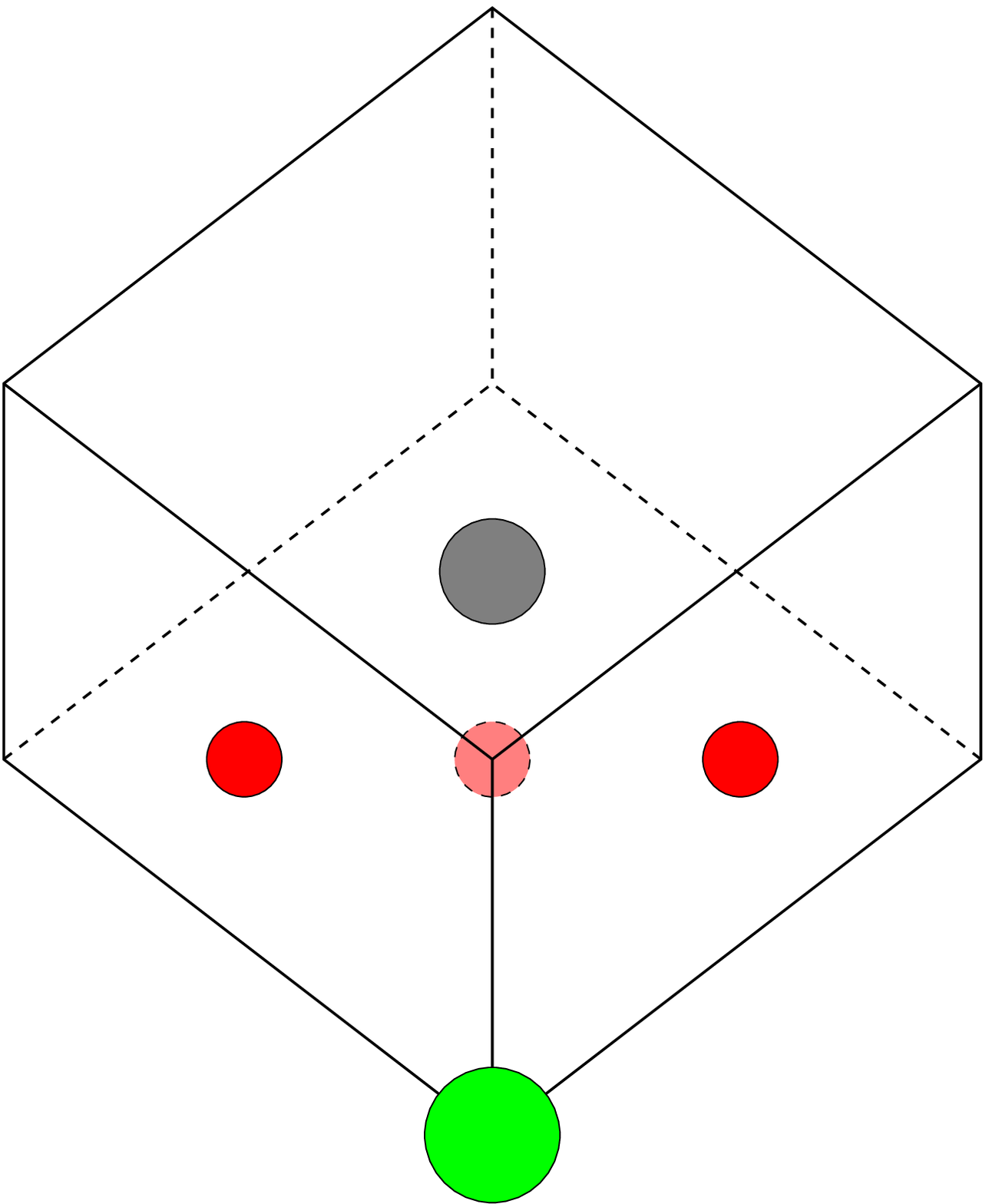} \hspace{0.5in}
\includegraphics[width=1in,clip]{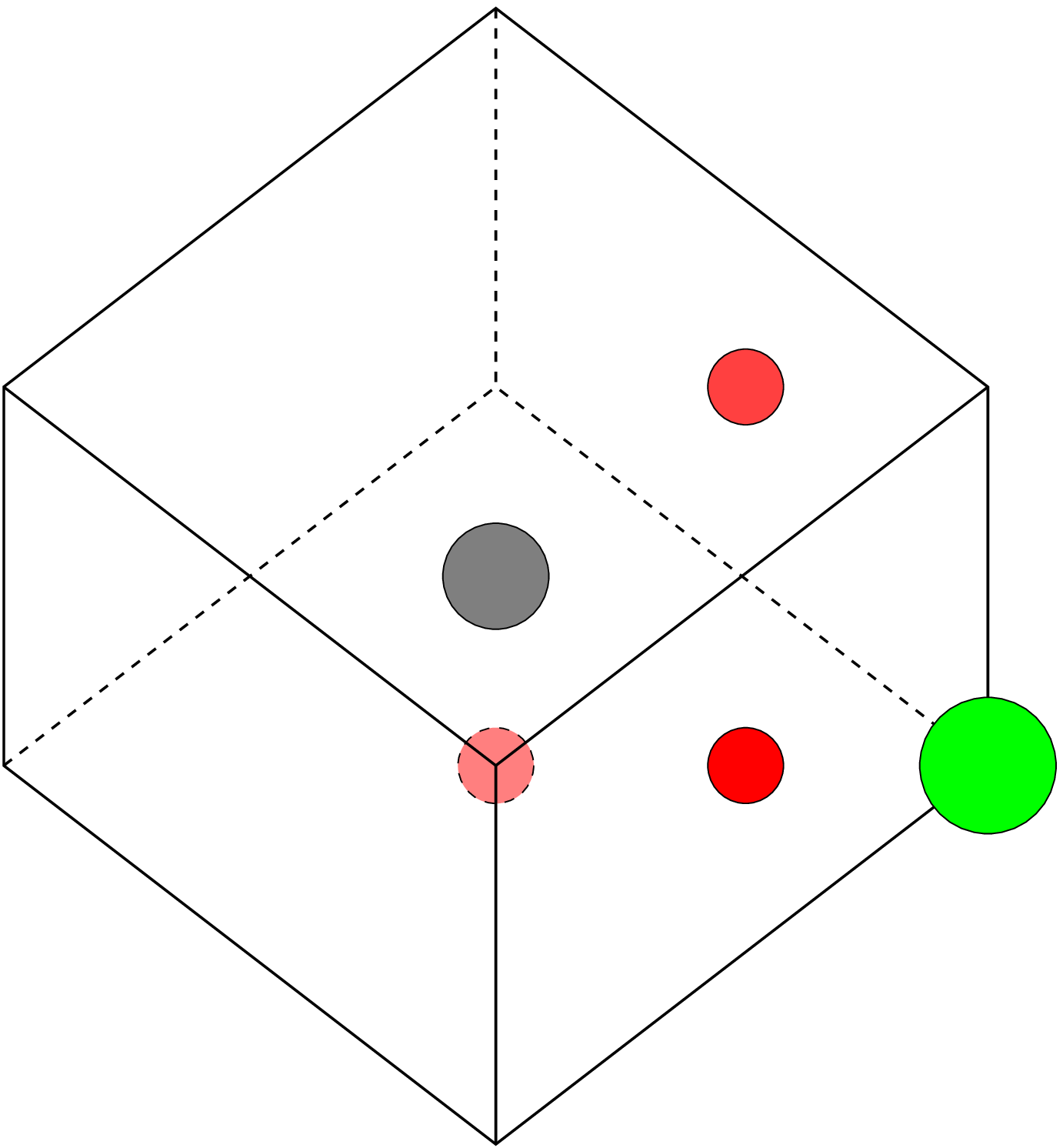}
\caption{ Primitive bulk SrTiO$_3$ unit cells for the model A (left) 
and B (right) (111) slabs. Sr atoms (large green circle) lie at the
corners of the cube; O atoms (small red circles) lie at the face-centered
sites; Ti atoms lie at the center of the cube. Both choices of the primitive
unit have zero dipole moment along any direction. \label{fig111cell} }
\end{center}
\end{figure}

We build two inequivalent stoichiometric slab models (that we call A and
B) by stacking the primitive building blocks schematically 
shown in Fig.~\ref{fig111cell}.
It is easy to verify that both choices of the primitive cell have 
zero dipole moment. (Again, to compute the dipole moment we use the 
\emph{formal} charges. This is substantiated by the Wannier-based 
decomposition described in section~\ref{sec:theory}, which provides the 
formal link to the theory of polarization in bulk 
solids.~\cite{King-Smith/Vanderbilt:1993}) 
The primitive translation vectors of the supercell are (in units 
of the bulk equilibrium lattice parameter $a_0=7.275$ a.u.)
${\bf a}_1 = (\sqrt{1/2},\sqrt{3/2},0)$, ${\bf a}_2 = (\sqrt{1/2},-\sqrt{3/2},0)$
and ${\bf a}_3 = (0,0,10)$; the out-of-plane spacing  
${\bf a}_3$ was chosen in order to include a sufficiently thick 
vacuum region separating the repeated images of the 10-layer slabs.
As the slabs do not enjoy inversion symmetry (there are a total of
four inequivalent surfaces in our simulations), we apply a dipole correction
in the vacuum layer to avoid unphysical macroscopic fields in the bulk
region of the SrTiO$_3$ films.
We use a regular $(8\times 8 \times 1)$ $\Gamma$-centered $k$-point mesh 
to sample the surface Brillouin zone, and we fully relax our structures
within the symmetry constraints allowed by the surface composition. Note
that the A-type slab preserves the point group of the bulk (111)
orientation, while the B-type slab has a lower symmetry due to the
presence of an incomplete oxygen plane on one side.

\begin{figure}
\begin{center}
\includegraphics[width=1.6in,clip]{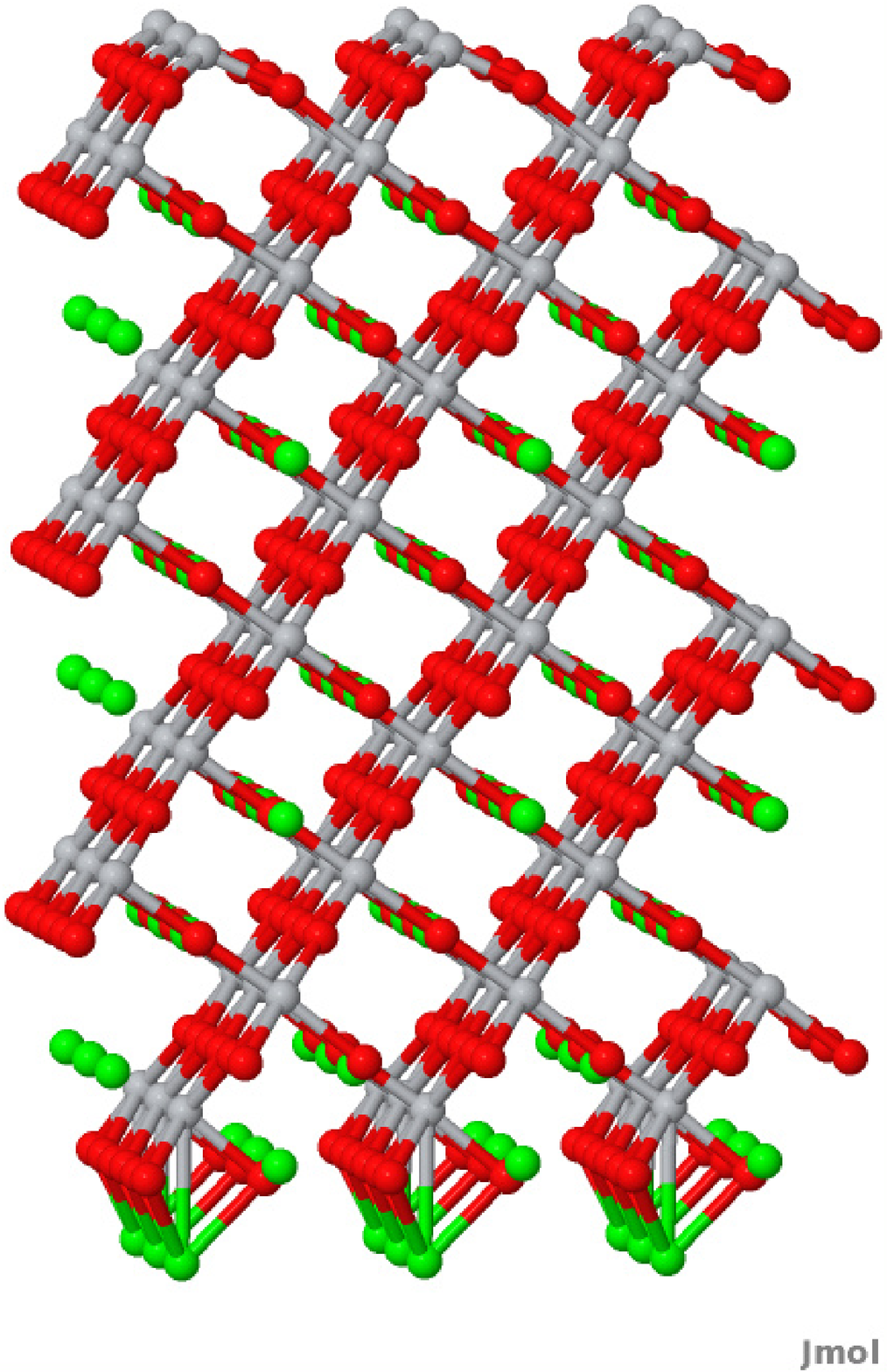}
\includegraphics[width=1.6in,clip]{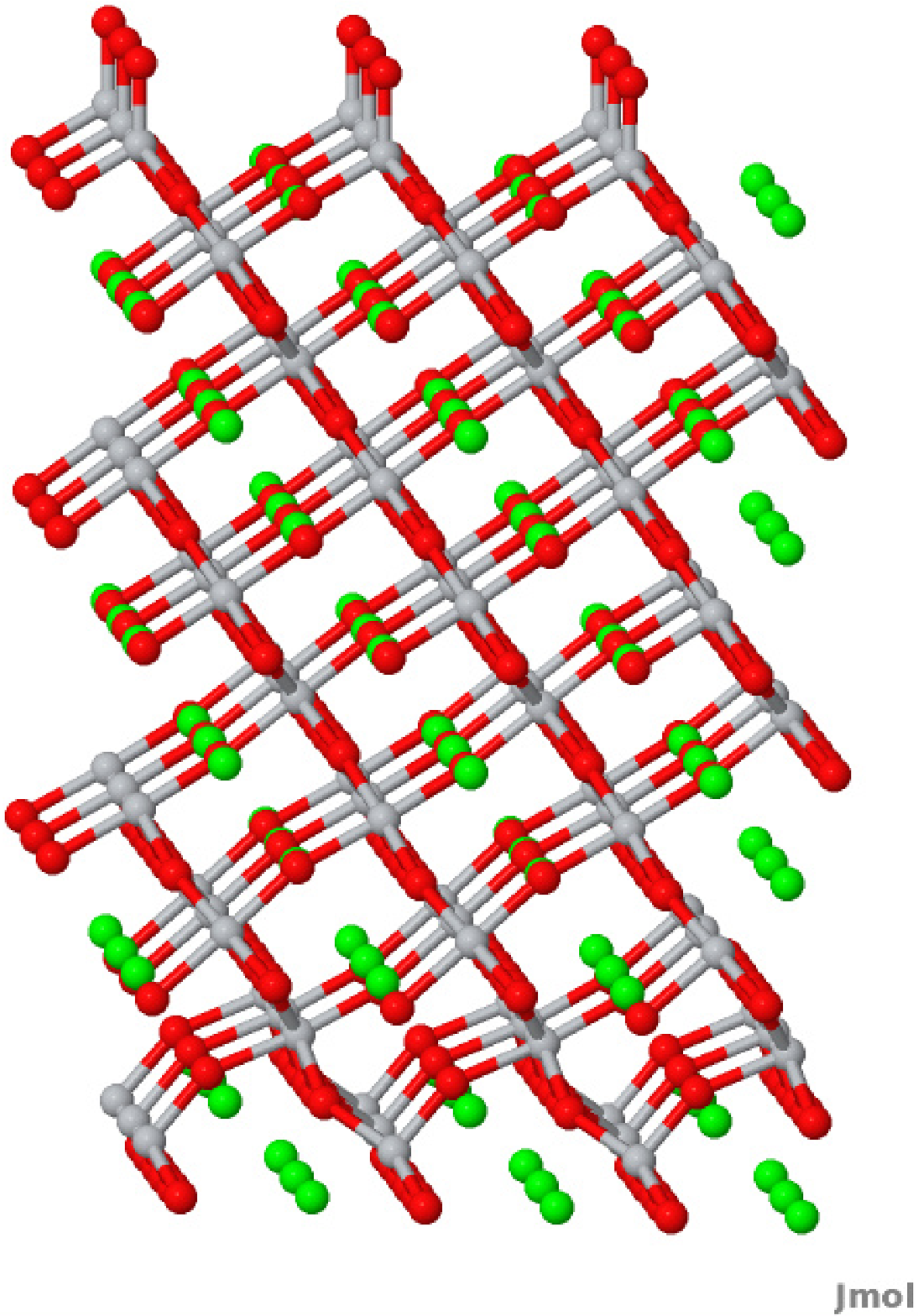}
\caption{ Relaxed geometries of the two (111) slabs described in the 
text. The left structure corresponds to model A; the right one to
model B. \label{fig:stoslab} }
\end{center}
\end{figure}

In Fig.~\ref{fig:stoslab} we show the relaxed structures of the two
slab models described above (the primitive unit of the supercell was
repeated three times in both in-plane directions to obtain a clearer
view of the structure).
Henceforth we shall indicate A1, A2, B1 and B2 the four inequivalent 
surfaces, where A and B refer to the specific slab model, and 1/2
refer to the top/bottom surface, respectively.
A1 has a TiO$_3$-type termination (i.e. an ideal Ti-type surface 
where the Sr atom has been removed from the topmost SrO$_3$ layer),
and correspondingly A2 contains a Sr-type termination, where 
undercoordinated Sr atoms protrude from the underlying oxygen group.
B1 has a supplementary O atom accommodated on top of an ideal Ti-terminated
surface, and this atom forms a tetrahedron surrounding the topmost Ti atom. 
B2 has an O vacancy in the terminating SrO$_2$ layer; model B can be therefore
obtained from model A simply by displacing a neutral SrO unit from the 2 (bottom)
to the 1 (top) surface.
Most of these surfaces were already considered in Ref.~\onlinecite{Blaha}, and 
indicated as ``small unit-cell reconstructions'' of SrTiO$_3$(111). 
We note that surface reconstructions are typically associated with a 
reduction in the translational symmetry group, which is not the case for
any of these models. Therefore, we rather regard these as primitive, 
stoichiometric bulk terminations.
Whatever is the nomenclature, the authors of Ref.~\onlinecite{Blaha} correctly
recognized the formal charge neutrality of these ``valence-compensated''
terminations.
 
\begin{figure}
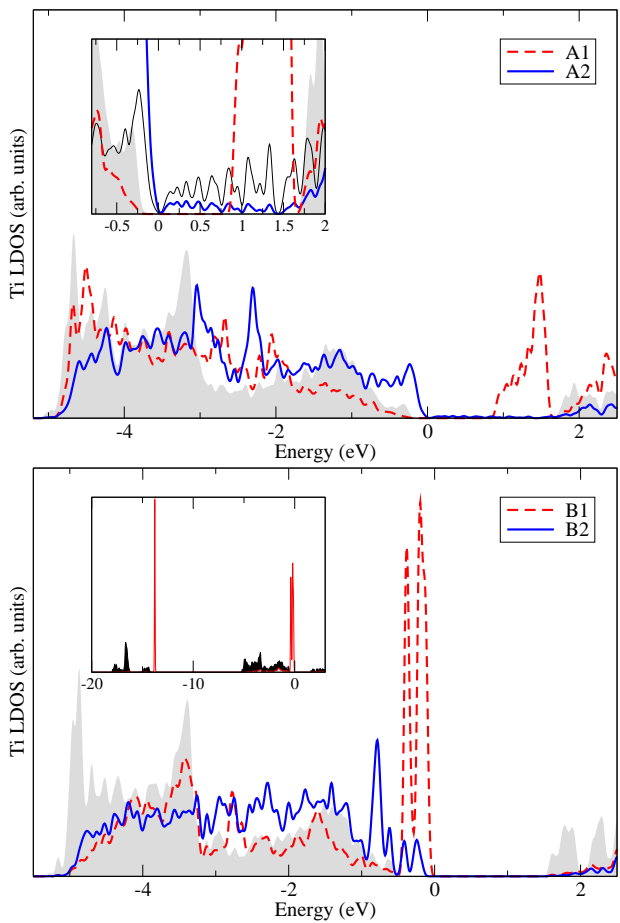

\begin{center}
\includegraphics[width=3.2in,clip]{fig8a.eps}
\includegraphics[width=3.2in,clip]{fig8b.eps}
\caption{ Local density of states (LDOS) on the
Ti atoms for the two slab models described in the text.
The Ti atom closest to top surface (type 1) corresponds to the 
red dashed curve; that lying closest to the bottom (type 2) surface
is indicated as a solid blue curve. The average LDOS on the two central 
Ti atom of either slab is shown as a shaded gray area. 
Top panel: model A; the inset shows a blow-up of the spectrum in a neighborhood 
of the Fermi level (the LDOS of the Sr atom closest to the bottom A2 
surface is also plotted as a thin solid curve). Bottom panel: model B;
the inset shows the atomic-like spectrum of the topmost O atom (solid
red curve with white shading), as compared with the bulk-like spectrum
of an O atom lying far from the surfaces (dark areas).
Note that in the A case (top panel) we used a finer $(16 \times 16 \times 1)$
$k$-point grid to compute the LDOS, in order to better describe the dispersive
surface state at the A2 termination.
\label{figldossto} }
\end{center}
\end{figure}

Similarly to the LaAlO$_3$ case, we analyze the electronic properties of
these surface models to verify their insulating character. We plot in 
Fig.~\ref{figldossto} the local density of states (LDOS) integrated on 
spheres of radius 3.0 bohr surrounding the Ti atoms. In the main panels 
we show the average Ti LDOS in the middle of the slab (gray shaded areas),
which we take as our bulk-like SrTiO$_3$ reference curve. We also show
the LDOS corresponding to the outermost Ti atom at the top (red dashed curve) 
and bottom (solid blue curve) surfaces. 
At A1 the gap is smaller than in the bulk, as a narrow band of Ti-derived
unoccupied orbitals splits from the conduction band. The band gap narrowing
is rather extreme at A2, where a highly dispersive surface 
state makes the gap as small as 0.1 eV at the $\Gamma$ point (presumably
this free-electron-like state is originated from the $s$ and $p$ state of 
the protruding Sr ions).
To better illustrate this, we show a blowup of the LDOS in the inset.
Here we also plot (thin black curve) the LDOS of the outermost Sr atom, 
where the surface state has its maximum weight. The nearly flat DOS (the
wiggles are caused by the finite $k$ resolution) between 0 and 1.5-2 eV, 
typical of a parabolic band in 2D, is clear.
The B slab presents, overall, an energy gap which is much closer to the
bulk value; this suggests that the system is electronically more stable than
in A.
Both at B1 and B2 the gap reduction is caused by valence-band derived surface
states; these are reminiscent of the states that are found at some BO$_2$-terminated
(100) perovskite surfaces. Conversely, no conduction band-derived states are present.
Especially interesting are the sharp peaks appearing at B1; these are derived from
the atomic-like orbitals of the outermost O atom. To illustrate this point, 
we plot in the inset the LDOS on the terminating O, which lies at the vertex of the
surface tetrahedron surrounding Ti; for comparison, we also show the LDOS of a bulk-like
oxygen far from the surfaces. 
The $2s$- and the $2p$- derived features of the surface O appear extremely sharp 
and atomic-like, in contrast with the substantial broadening in the SrTiO$_3$ 
bulk caused by band dispersion. 
A tetrahedral coordination might look unusual for Ti, which tends to adopt 
octahedral coordination in most (if not all) stable bulk oxide phases. 
Nevertheless, at the SrTiO$_3$ surface, analogous tetrahedral units 
were recently shown both experimentally and theoretically to be 
energetically favorable,~\cite{Enterkin-07} even in the case of the
(110) orientation where there exist alternative $(1\times 1)$ structures
with relatively low energy.~\cite{Eglitis:2008}

Finally, we shall comment on the relative energy of these structures. Unlike the
$(n10)$ models discussed in the previous section, here it is not possible to 
construct a stoichiometric \emph{and} symmetric slab; therefore, we can only 
calculate a \emph{cleavage} energy for A and B, $E_{\rm cl}$. Resolving this 
value into the contributions of the top and bottom terminations would require 
further considerations about the chemical potential of Sr, Ti and O; this goes
beyond the scopes of the present work. 
%
%
%
%
We find $E_{\rm cl}$(A)= 6.27 J/m$^2$ and $E_{\rm cl}$(B)= 3.94 J/m$^2$. These
values are both larger than the previously reported cleavage energies 
along the (100) or (110) directions. Especially the A model has a high energy
cost, consistent with the ``open'' nature of the low-coordinated surface sites,
and with the relatively unfavorable electronic configuration discussed in the 
previous section. Model B, on the other
hand, has a cleavage energy that is significantly smaller, and (on average) 
reasonably close to typical (110) surface energies.
It should be kept in mind that the cleavage energy might be unequally distributed 
between the B1 and B2 surfaces -- we cannot exclude that one of the two might be 
quite stable in a wide range of thermodynamic conditions.
$E_{\rm cl}$(B) can be directly compared to the values reported in Ref.~\onlinecite{Blaha},
where it appears to correspond to the sum of the surface energies of model 3 and 4.
The authors of Ref.~\onlinecite{Blaha} did not use LDA but a variety of different density
functionals, with values ranging from 4.94 eV (PBE) to 6.41 eV (TPSSh) per surface 
cell. Our LDA value of 6.31 eV per surface cell compares favorably with the highest
value reported there, consistent with the systematic tendency of LDA towards 
overbinding. 

It is interesting to compare our calculated $E_{\rm cl}$(A)=10.0 eV/cell to the 
energy associated with the ``textbook'' cleavage, i.e. that leaving atomically flat, 
metallic and polar Ti / SrO$_3$ terminations. 
Assuming that our LDA values are comparable to the TPSSh results of Marks {\em et al.},
we can infer the Ti / SrO$_3$ cleavage energy by summing up the TPSSh surface energies of 
model 1 and 2 in the aforementioned work; this yields a value of 12.3 eV / cell.
As surprising as it may sound, our non-polar cleavage model A, with the severely 
undercoordinated Sr atoms protruding from surface A2, is still about 2 eV /cell lower
in energy than the atomically flat cleavage model. This fact highlights the
importance of achieving electrostatic stability from bulk-like building blocks,
without invoking external compensation mechanisms (such as hole or electron doping 
as in the case of Ti / SrO$_3$).

\subsection{Polarity compensation of LaAlO$_3$(100)}

\begin{figure}
\begin{center}
\includegraphics[width=3.0in,clip]{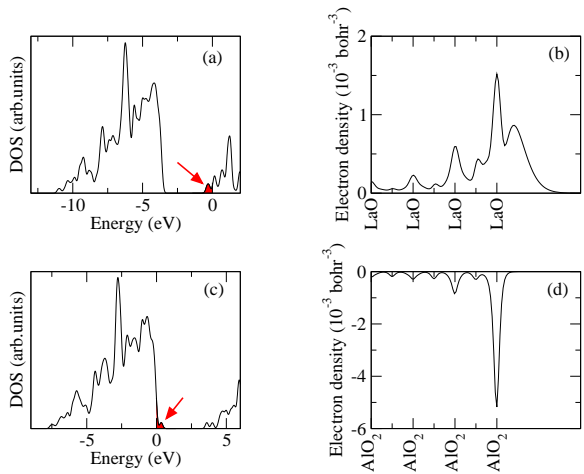}
\caption{ (a) and (c): Total DOS of the (100) LaAlO$_3$ slabs, highlighting
the population of the states involved in the compensation of the surface
polarity (shaded area, indicated with an arrow). (b) and (d): Plane-averaged
density of compensating charge, related to the shaded portion of the DOS in
(a) and (c). Top (a-b) and bottom (c-d) panels refer to the LaO- and 
AlO$_2$-terminated slabs, respectively.
\label{figmetal} }
\end{center}
\end{figure}

So far we restricted our analysis to ideal non-polar surfaces, 
i.e. systems where only bulk-like building blocks are present.
To complete our discussion we now consider a prototypical 
polar surface, LaAlO$_3$(100), and illustrate how our arguments 
apply to the analysis of selected compensation mechanisms,
where we introduce extrinsic sources of compensating charge $\sigma_{\rm ext}$.

\subsubsection{Via metallic carriers}

First, we consider the clean $(1 \times 1)$ LaO- and AlO$_2$-terminated
(100) surfaces. Due to the built-in dipole of the bulk unit cell that we
must use to construct these terminations, there is an excess charge of
$+0.5e$ and $-0.5e$ per surface unit cell, respectively. If we don't relax
the translational symmetry, the only possible compensation comes from 
metallic carriers, either in the form of conduction band electrons or 
valence holes. 
By performing two separate calculations of symmetrically terminated
6.5-unit cell thick slabs, we indeed obtain metallic surfaces. In 
Fig.~\ref{figmetal} we plot the total density of states for both
slabs, where the Fermi level clearly crosses either the valence band 
(AlO$_2$ termination) or the conduction band (LaO termination).

An interesting feature of the DOS of Fig.~\ref{figmetal}(a-c) is that, in
both cases, a clear gap persists in the spectrum. This means that, in 
spite of the partial metallization, the conduction and valence bands 
preserve their respective identities.
This observation implies that we can rigorously separate 
what we consider ``bound charges'' (which are all of bulk 
origin here, as we don't introduce extrinsic species in the supercell) 
from ``external compensating charge'', following the prescriptions 
of Ref.~\onlinecite{Band} and~\onlinecite{laosto-11}.
The former, which we take as the total charge density of the 
(completely filled) valence-band manifold, are implicitly included
in the definition of ${\bf P}_{\rm bulk}$; the latter can be either 
a positive external density of valence-band holes (ext,h) or a 
negative density of conduction-band electrons (ext,e), 
%
\begin{eqnarray}
\rho_{\rm ext,e}(z) & = & \int_{E_{\rm mid-gap}}^{E_{\rm F}} \tilde{\rho}(E,z) dE \\
\label{eqrhoe}
\rho_{\rm ext,h}(z) & = & -\int_{E_{\rm F}}^{E_{\rm mid-gap}} \tilde{\rho}(E,z) dE.
\label{eqrhoh}
\end{eqnarray}
Here $\tilde{\rho}(E,z)$ is the planar-averaged and energy-smeared local
density of states defined in Ref.~\onlinecite{Band}, and $E_{\rm F}$ is the
Fermi level. The electronic states that contribute to the integrated charge
densities $\rho_{\rm ext,e}$ and $\rho_{\rm ext,h}$ are evidenced as 
shaded areas in the DOS plot of Fig.~\ref{figmetal}. (Note that the DOS is 
the volume integral of $\tilde{\rho}(E,z)$.) 
In Fig.~\ref{figmetal}(b-d) we plot the compensating surface densities 
$\rho_{\rm ext,e}$ and $\rho_{\rm ext,h}$. Both appear localized to 
the surface region, although they display a relatively slow decay into
bulk LaAlO$_3$, and amount (within machine precision) to a total 
of exactly plus or minus half an electron per side. This demonstrates 
the full consistency [in the sense of Eq.~(\ref{eq:neutral})] between 
the ``external charge'' defined in Eq.~(\ref{eqrhoe}) and~(\ref{eqrhoh}), and the 
prediction of excess bound charge coming from the analysis of 
${\bf P}_{\rm bulk}$.
Note that, in the case of the LaO-terminated slab, part of the charge 
spills out into the vacuum region. This is a consequence of the vacuum
level being very close to the conduction band edge, which is  
populated by the compensating electrons. Conversely, only O($2p$)-derived
states contribute to $\rho_{\rm ext,h}$.
By combining the total energies of the reference slabs and 
subtracting an appropriate number of bulk reference units,
we obtain a relaxed cleavage energy of 4.53 eV per surface 
cell (5.13 J/m$^2$). This is larger than the cleavage energies 
we computed in Sec.~\ref{sec:vicinal} for the primitive non-polar 
$(n10)$ surface models.

\subsubsection{Via external bound charges}

\begin{figure}
\begin{center}
\includegraphics[width=3.0in,clip]{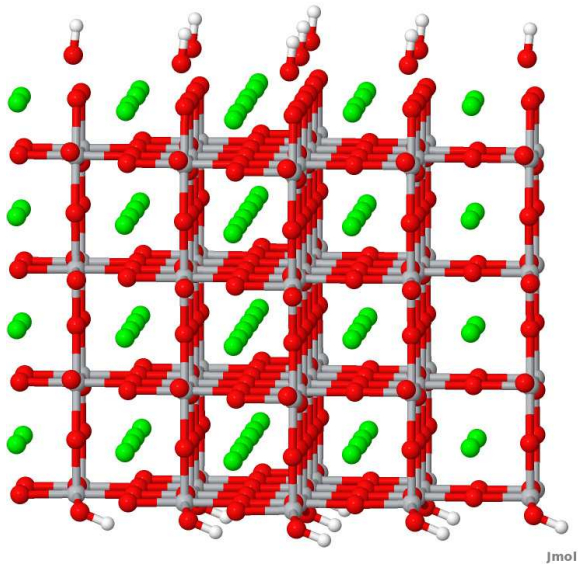}
\caption{ Relaxed structure of the LaAlO$_3$(100) slab with
LaO (top) and AlO$_2$ (bottom) terminations, compensated 
with OH(-) and H(+) groups respectively. 
\label{figh2o} }
\end{center}
\end{figure}

We shall now consider a different compensation mechanism, where instead 
of metallic carriers the surface acquires bound charge via adsorption of 
external species.
As a matter of fact, this is of concrete relevance for the interpretation
of a vast number of physical phenomena. A real surface is always in 
contact with an atmosphere where various gas-phase species are present, 
and a number of exchange/adsorption/redox processes are usually
thermodynamically accessible.
%
On a more general basis, there are many situations where the surface layer 
differs, either compositionally and chemically, from the bulk of the crystal.
Consider, for instance, the Sr-decorated Si(100) surface that is used to 
promote coherent epitaxial growth of SrTiO$_3$.~\cite{Blochl-04,Kolpak-10}
Given their technological and fundamental importance, it is useful here to 
provide some examples, without the pretention to be exhaustive, of how our 
arguments can be translated to address those situations.

In the specific case of LaAlO$_3$, H atoms adsorbed at the surface
of a film deposited on a SrTiO$_3$ substrate were found to significantly
alter the electrical boundary conditions, e.g. by reducing or enhancing the 
residual internal electric field in LaAlO$_3$ and by influencing the free
carrier concentration at the interface.~\cite{Son-10} Experimentally,
a humid atmosphere was demonstrated to be necessary to stabilize conducting
paths at the buried interface~\cite{watercycle}. In turn, these ``writing''
and ``erasing'' processes~\cite{Cen-08} appear to be mediated by charged 
surface adsorbates~\cite{chargewrite}.
All in all, there is growing evidence that OH and H species are crucial
to explain many outstanding phenomena experimentally observed at LaAlO$_3$ 
surfaces and thin films. Hence the motivation for studying H$_2$O-based
compensation mechanisms, where the surface retains the insulating character 
of the bulk.

Adding external species to a $(1 \times 1)$ surface won't make it insulating,
as the excess charge is half an electron per cell (a periodic array of
external species can provide only integer multiples of $e$). Doubling the surface 
unit cell leads to exactly plus or minus one electron, that now allows for
an insulating state. We compensate this excess charge with a
split water molecule (H adsorbed on the ``negative'' AlO$_2$ side, OH on the 
``positive'' LaO side) per $\sqrt{2} \times \sqrt{2}$ surface cell.
We use a stoichiometric LaAlO$_3$ slab with a thickness of four unit cells,
a $c(2\times 2)$ in-plane translational symmetry and we relax the structure
without imposing any symmetry constraint. As usual, we use a vacuum dipole 
correction to ensure the correct cancellation of the macroscopic electric fields
due to the asymmetry of the slab.
At equilibrium, the structure appears as in Fig.~\ref{figh2o}. Note the tilted
position of the H atoms on the AlO$_2$ side, consistent with the geometry 
found in Ref.~\onlinecite{Son-10} for the H-LaAlO$_3$/SrTiO$_3$ system. The OH groups
on the LaO side lie in a bridge site between two surface La atoms, thus occupying
a natural lattice site for O. (An analogous location of the OH group was 
found on the SrO-terminated SrTiO$_3$ surface decorated with dissociated
water~\cite{Guhl-10}.)
The system has a large insulating gap, almost equal to the bulk value,
suggesting that this configuration might be fairly stable. We can estimate
the energetics by considering a ``wet cleavage'' experiment where two
LaAlO$_3$ (001) surface are created and at the same time one free H$_2$O 
molecule is split between the two terminations,
\begin{equation}
E_{\rm cl} = E_{\rm slab} - 8 E_{\rm bulk} - E_{\rm H_2O}.
\end{equation}
Here $E_{\rm slab}$ is the energy of the supercell described above with 
two LaAlO$_3$ cells per surface unit and a thickness of four unit cells;
$E_{\rm bulk}$ is the bulk energy, calculated by including the
antiferrodistortive tilt of the O octahedra, which now are allowed by symmetry;
$E_{\rm H_2O}$ is the energy of a free water molecule, calculated by using a
cubic box of approximately 10 \AA{} lateral size.
The resulting cleavage energy per surface area
is 2.50 J/m$^2$, which is the lowest value calculated in this work.
This result suggests that adsorption of OH and H groups, which are 
ubiquitous in most experimental setups, is a very likely 
candidate to stabilize the LaAlO$_3$ surface polarity.
A study of LaAlO$_3$(100) compensation via point defects was also 
recently reported in Ref.~\onlinecite{Demkov-11}.

As a final remark, note that bound compensating charges, unlike
the metallic carriers mediating electronic compensation, come in 
discrete units of $e$. Therefore, in cases where bound-charge compensation
occur, it is most appropriate to ``count'' the external charges per
unit area, which must satisfy the relationship
\begin{equation}
{\bf P}_{\rm bulk} \cdot \hat{n} = -\frac{Q}{S}.
\label{qs}
\end{equation}
Here $Q=ne$ (with $n$ integer) is the formal oxidation state of the
external defect or adsorbate, and $S$ is the surface area per defect.
Note that there exist defects (e.g. transition metal cations) that
are stable in several oxidation states; of course, the actual $Q$ that
occurs in the situation of interest must be used in Eq.~(\ref{qs}). 
In case of doubt, the Wannier-based analysis of Sec.~\ref{sec:theory}
can be used to assess the formal oxidation state of a given defect.

\section{Discussion}

\label{sec:discussion}

Here we shall put our results in the context of the current
state-of-the-art in the field, especially regarding the fundamental 
theoretical understanding of the electrostatic stability of insulator 
surfaces.

\subsection{Insulating nature of the surface}

As we mentioned several times when discussing our applications, it
is likely that a non-polar (in the sense specified in this work)
primitive surface will have a well-defined surface band gap. 
Here we shall further specify this point, to prevent dangerous
generalizations. 

It is certainly true that a polar surface with
all the atoms in their bulk oxidation state cannot exist. If
we insist on keeping the local stoichiometry fixed, some of the
atoms must change their valence in order to avoid a diverging
electrostatic energy. In many cases this produces partially
filled electronic bands and a metallic surface. 
It is not difficult to imagine cases, however, where the surface 
atoms may change their oxidation state while preserving a gap in 
the spectrum. This would happen, for example, whenever the 
excess/defect charge amounts to an integer number of electrons, 
and there are ions in the lattice that have multiple stable 
oxidation states, e.g. most transition metals.
Oxygen might also, in principle, change its formal valence from
-2 to -1 to compensate a net surface charge -- such a mechanism,
stabilized via the formation of a peroxo bond, was reported in
the case of a SrTiO$_3$ surface by Bottin {\em et al.}~\cite{Bottin-03}.
Therefore, a polar surface does not necessarily lead to a metallic
surface. There are several other compensation mechanisms available
(often accompanied by a reduction in the translational periodicity)
that leave the surface insulating even without changes in the 
stoichiometry. We stress that in this latter case, however, the 
formal oxidation state of some atoms \emph{must} change.

Also the statement that non-polar surfaces are insulating is
far from universal. Indeed, by replicating a sufficiently pathological
choice for the bulk primitive unit, one might end up with a surface 
that has a very awkward bonding configuration. This might produce a
dramatic departure from the bulk bonding environment, and in such cases 
it is well possible that one or more surface bands may close the band
gap. An example of how this may happen is provided by
our A2-type SrTiO$_3$ surface, where the band gap is reduced to a 
tiny value of about 0.1 eV. Gap closure would also occur 
for our $(n10)$ LaAlO$_3$ surface models for a large enough $n$; 
at some point the electrostatic energy of the large LaO- and AlO$_2$-type
terraces would become too large and eventually the gap would close.
Note that surface relaxation usually helps stabilizing the 
truncated bonding network; in cases where our arguments would
predict an insulating and non-polar surface, it is not infrequent 
to observe that a sizeable band gap opens only after full atomic 
relaxation.

In conclusion, the relationship between electrostatic stability
and insulating nature is certainly not a rigorous one. It is nonetheless
a useful guideline, in the sense that if the surface bonding environment
is not too pathological and the solid has a marked ionic character one 
usually expects a non-polar surface to be insulating.
Note that the insulating/metallic nature of a formally non-polar termination
is more an issue of chemistry than of electrostatics -- it boils down
to the chemical driving force for the atoms to preserve their bulk-like
oxidation state. If the surface becomes metallic this will 
happen through a \emph{local} rearrangement of the electron cloud;
the total surface charge won't change.

\subsection{Covalency arguments and ``weak polarity''}

The (001) surfaces of II-IV perovskites such as SrTiO$_3$, BaTiO$_3$, 
etc. are classified as non-polar within our definitions, consistent
with Tasker's assumption of formal ionic valence.
Goniakowski, Finocchi and Noguera~\cite{Noguera-08} challenged 
Tasker's classification by invoking covalent bonding effects, which 
would produce a smearing of the electron cloud. This, in turn, would 
produce a distribution of the charge between the oxide layers that 
differs from their formal ionic charges. 
According to this interpretation, SrTiO$_3$(001) was classified as
``weakly polar''. 

The disagreement between the two interpretations is rooted in the 
way the electronic charge density is partitioned into individual 
building blocks at the bulk level. 
Choosing Mulliken or Bader populations inevitably leads to a 
fractional charge per ion that is typically smaller than the 
(integer) formal oxidation state, and the individual SrO and 
TiO$_2$ layers appear charged. 
%
%
While these ideas have certainly some merit, there are severe drawbacks
as well. The most important one is that, by insisting on a  
partition of the electron cloud based exclusively on the total 
charge density $\rho({\bf r})$, one thwarts any further attempt at 
linking the discussion of surface electrostatics to the theory of 
polarization in bulk solids. 
Powerful and rigorous results of the modern theory of polarization,
e.g. the ``interface theorem'', are inconsistent with a description
of surface polarity in terms of Mulliken,
Bader or even Born effective charges. 
(The fact that in II-IV perovskites the individual AO and BO$_2$ layer do 
not satisfy the acoustic sume rule separately is sometimes taken as a 
further argument in support of the weak polarity concept.)
The theory of bulk polarization 
implies a \emph{wavefunction}-based partition of $\rho({\bf r})$. 
The natural tool in that sense are the spatially localized Wannier 
function as we discussed in section~\ref{sec:theory}. 
Covalency effects are irrelevant in this context, 
except that they are implicitly accounted for through the location 
of the ground-state Wannier centers and their spatial spread.
Using Wannier functions might appear unnatural and complicated at first 
sight, but eventually they really lead to drastic simplifications and to 
an intuitive physical picture.
In particular, if one wants to recover the intuitive classical formula 
${\bf P} \cdot \hat{n} = \sigma$, there is simply no other way around.

We believe that full consistency between the theory of bulk polarization
and the theory of surface polarity is a must. Therefore, we caution
against the use of concepts such as weak polarity or covalent charges as
they are are inconsistent with the former.

\subsection{Other oxide surfaces}

\subsubsection{Ferroelectric perovskites}

As the concept of surface polarity is intimately linked to the
polarization of the bulk solid, ${\bf P}_{\rm bulk}$, it is particularly
insightful to discuss cases where ${\bf P}_{\rm bulk}$ has a non-trivial
behavior, as in ferroelectric perovskite materials.
Consider a (100)-oriented slab of BaTiO$_3$, with the spontaneous 
polarization vector, ${\bf P}_{\rm S}$ oriented along the normal to 
the surface. 
Imagine that we have a stoichiometric slab with ideal BaO and TiO$_2$
terminations, and a monodomain state with perfect $1\times 1$ periodicity;
assume also that ${\bf P}$ points towards the TiO$_2$-type surface.

Both surfaces are polar, as ${\bf P}_{\rm bulk} \cdot \hat n = P_{\rm S} \neq 0$.
Contrary to the LaAlO$_3$ example, however, here ${\bf P}_{\rm bulk} \cdot \hat n$ 
is not a simple fraction (plus or minus one half) of the polarization quantum $e/S$. 
Here $P_{\rm S} = pe/S$, where $p$ is a real number of the order of 0.25-0.35 (depending
on the in-plane strain imposed to the film). Therefore, it might be technically 
difficult in a calculation to construct a commensurate supercell where ${\bf P}_{\rm S}$
is accurately compensated by an appropriate coverage of charged 
adsorbates or defects (unless $p$ happens by accident to be exactly equal to a 
rational number with a small denominator).
A possible trick to circumvent this difficulty is using the so-called ``virtual crystal 
approximation'' (VCA)~\cite{Band}. Here a fractionally charged pseudopotential is introduced
at the surface to reproduce the effect of a disordered array of defects with the appropriate 
coverage. This way, the surface can be made insulating and charge-neutral at a low 
computational cost; the price to pay is that the VCA does not lend itself easily to the
calculation of surface-specific properties, e.g. the energetics of a given compensation 
mechanism.

A second important example is that of I-V ferroelectric perovskites, e.g. KNbO$_3$.
Here the ferroelectric contribution to the polarization, $P_{\rm S}$, adds
up to the ``compositional'' built-in cell dipole,~\cite{laosto-11} which is 
$P_0 = \pm e / 2S$ (as in LaAlO$_3$, the layers are formally charged, 
although here AO layers are negative and BO$_2$ are positive).
Note that, if $|P_{\rm S}|$ were (again, by accident) equal to half a quantum
of polarization, both NbO$_2$- and KO-terminated (100) surfaces would be non-polar
(i.e. $P_{\rm S}$ would cancel out $P_0$), provided that the spontaneous polarization 
points in the correct direction.
This would be away from the surface for the $p$-type NbO$_2$ termination, and
towards the surface for the KO termination. 

Finally, it is worthwile mentioning the case of BiFeO$_3$. This material appears 
complicated at first sight, because of the tilted polarization axis 
[${\bf P}_{\rm S}$ is oriented along the (111) direction] and the compositional 
layer charges of $\pm 1$ (both Fe and Bi are formally 3+ ions). 
However, within the formalism established in this 
work, predicting the excess surface charge density that will be present at a given 
ideal termination becomes trivially simple.
For example, at the FeO$_2$ (100) termination we have [just like in the case of the 
AlO$_2$-terminated LaAlO$_3$(100) surface] an excess built-in charge of 
$-e/2S \sim -0.5$ C/m$^2$. To this value we need to add the 
projection of ${\bf P}_{\rm S}$ along the (100) axis, which amounts to approximately
the same value.~\cite{Neaton-05} Therefore, if ${\bf P}_{\rm S}$ points towards
the FeO$_2$-type (100) termination, this surface will be in practice only very weakly
charged or even neutral. 
It goes without saying that the composition of the surface ``pins'' the out-of-plane
component of the polarization to a fixed value that cannot be switched (unless
the surface composition itself is changed, see Ref.~\onlinecite{chemswitch}).

\subsubsection{ZnO}

The use of ZnO in many technological areas, as well as the recent progress in 
fabricating tailored nanostructures and functional surfaces with this material,
have generated a widespread interest in the fundamental properties of its
polar (0001) surface.~\cite{Wander-01,Valtiner-09,Lauritsen-11}.
Several possible compensation mechanisms involving, e.g. metallic free carriers~\cite{Wander-01},
hydroxylation/protonation~\cite{Valtiner-09} or stoichiometry changes~\cite{Lauritsen-11}
have been proposed over the years.
In spite of this activity, the question of exactly {\em how much} excess charge is 
present at the polar Zn- or O- terminated surfaces is still a source of confusion.

For instance, there is a common belief that, starting from an ideal unreconstructed
termination, removal of 1/4 of the surface ions will lead to perfect compensation
of the polarity.~\cite{Lauritsen-11} 
This would be true if ZnO crystallized in zincblende phase. However, bulk ZnO is 
wurtzite-type, which means that on top of the compositional
(zincblende-like) dipole it has also a non-trivial spontaneous $P_{\rm S}$.~\cite{DalCorso-94}
This $P_{\rm S}$ is of course not switchable, unlike the ferroelectric materials 
discussed in the previous section, but it does need to be taken into account when 
computing the surface charge.
First-principles calculations of $P_{\rm S}$ have reported relatively small
values (compared to a hypothetical zincblende reference structure) of $P_{\rm S}$
in bulk ZnO, of the order of 0.02-0.07 C/m$^2$~\cite{Priya,DalCorso-94}.
This implies that the necessary correction to the zincblende-like excess charge
of $0.5e$ per surface cell is of the order of 0.01-0.03 electrons. Even if this 
correction is not large, one should keep in mind that, in a hypothetical free-slab
calculation of ZnO where 1/4 of the O (and Zn) surface ions have been removed,
after full relaxation there will be a non-zero residual macroscopic electric 
field in the slab of approximately $\mathcal{E}_{\rm slab} = P_{\rm S} / 
(\epsilon_0 \epsilon_{\rm r})$. Here $\epsilon_0$ is the vacuum permittivity
and $\epsilon_{\rm r}$ is the static dielectric constant of ZnO (including
piezoelectric effects). In fact, this observation was used to calculate the
spontaneous polarization of wurtzite BeO several years before the modern
theory of polarization was developed.~\cite{Posternak-90}

\subsection{Semiconductor surfaces}

While our arguments apply most naturally to ionic materials, where the
assignment of the localized Wannier charges to a given atom is unambiguous,
with some care they can be easily adapted to covalently bonded insulators.
The main difficulty is that in semiconductors (e.g. Si) the maximally-localized 
Wannier functions tend to occupy bond-centered sites, and are shared between two
atoms -- assigning a given Wannier function to either atom that 
participate to the bond is then entirely arbitrary. 
Nevertheless, one can usually establish a reasonable convention for partitioning 
the bulk solid into well-defined units. For instance, in Si one could assign
four spin-up Wannier functions (their centers would form a tetrahedron around
the nucleus) to one atom and four spin-down Wannier functions to the other atom
in the basis. (It might appear somewhat artificial to use such a spin-split basis;
however, for the present discussion, the information about the spin is 
irrelevant, only the charge density of the Wannier functions really matters.) Then,
this decomposition yields a basis of two WIs that individually retain the full
symmetry of the lattice, are charge-neutral and have zero dipole moment. 
Primitive Si surfaces are then predicted to be non-polar, but chemically they 
will be highly reactive because of the singly occupied ``dangling bonds''; this
picture is consistent with the widely accepted understanding of Si surfaces.
It is easy to see that by saturating these bonds with H one always obtains a 
non-polar and chemically stable surface (H does not add a net charge density 
as it contributes one electron and one proton to each dangling orbital).
Alternatively, one could supply one Sr atom every two dangling bonds; this
stabilization mechanism is important for growth of perovskite oxide films
on Si substrates.~\cite{Blochl-04,Kolpak-10}
Interestingly, in the case of the Sr-decorated surface, further oxidation 
does not change the surface charge count,~\cite{Blochl-04} as additional 
O atoms achieve a closed-shell configuration by incorporating the electron 
pairs  already present in the saturated dangling bonds.
This is a system where oxygen adsorption does not change the surface charge,
in striking contrast with typical ferroelectric surfaces.~\cite{chemswitch}

Of course, one could prefer to use other conventions, e.g. assign two doubly 
occupied Wannier functions to each Si atom. This way the Si(100) or (111) surfaces would be
understood as ``polar'', and they \emph{are} indeed polar if one insists on counting
electrons two-by-two (the dangling bonds would need to be either empty or saturated,
without the necessary countercharge to balance the electrostatics).
This means that the concept of polar surface becomes somewhat 
ill-defined if the solid has no ionic character whatsoever. Note that 
the formalism developed in Ref.~\onlinecite{Vanderbilt/King-Smith:1994},
on which the present work heavily relies, provides always a rigorous 
means of calculating the surface charge from bulk properties, 
regardless of the (ionic or non-ionic) nature of the insulator,
and independently of the convention that one uses to ``assign'' 
the bound electron charges to a given lattice site.
A more extensive treatment of the covalent case can be found in
Ref.~\onlinecite{Vanderbilt/King-Smith:1994}, and was recently
discussed also in Ref.~\onlinecite{Bristowe-11}.

\subsection{Interfaces}

In this work we decided to focus on surfaces, which is a special case of
interface between two materials (one of them is vacuum). Whenever
the second material is another crystalline insulator, the same arguments apply,
but the ``electrostatic phase diagram'' can be substantially richer.
The simplest case is that of two materials that have the same crystal structure,
and we assume coherent epitaxy, i.e. both semi-infinite regions have the same 
in-plane periodicity and the same crystallographic orientation of the atomic planes.
However, there might be more complex cases -- for example, the participating materials
have different bulk structures, or they are not oriented along the same crystallographic
direction.
In any case, the electrostatics is always governed by the intuitive classical
formula,
\begin{equation}
({\bf P}_2-{\bf P}_1) \cdot \hat n = \sigma_{\rm ext}.
\end{equation}
Here ${\bf P}_{1,2}$ is the polarization in either material, calculated by choosing a 
certain basis for the primitive basis of atoms and Wannier functions;
$\sigma_{\rm ext}$ is the ``remainder'' interface charge, that is left behind
once one removes all the bulk-like primitive units on either side;
$\hat n$ is the normal to the surface plane.
As in the case of surfaces, we define an interface non-polar if, for an ideal
termination of both materials with the maximum allowed translational symmetry
one has $\sigma_{\rm ext}=0$.

\section{Conclusions}

\label{sec:conclusions}

In summary, we have revisited the concept of polar surface within the
context of the modern theory of bulk polarization. Our definition, which
is consistent with the bound (and discrete) nature of electrons in the 
insulating state of matter, puts Tasker's classification on a firmer theoretical 
grounds, and corroborates it at the microscopic level. We further complete
Tasker's formalism with an additional term, which comes from the polarization
of the electron cloud in solids that spontaneously break space inversion symmetry.
Our calculations of non-polar LaAlO$_3(n10)$ and SrTiO$_3$(111) surfaces, and of 
compensation mechanisms at LaAlO$_3(100)$, demonstrate that our formalism provides
a convenient way of describing the net surface charge in terms of bulk polarization 
and external sources (either ``bound'' or ``free'').
We have also illustrated some practical analysis tools that can be used to monitor 
the equilibrium distribution of compensating charge in a calculation.
We hope that these techniques will be helpful for future first-principles studies,
and more generally as a conceptual basis to rationalize the many interesting phenomena 
occurring at the surfaces of insulating materials.

\section*{Acknowledgments}

I am indebted to D. Vanderbilt for a critical read of the manuscript
and many illuminating discussions.
This work was supported 
 by DGI-Spain through Grants No. MAT2010-18113 and No. CSD2007-00041,
 and by the European Union through the project EC-FP7, 
 Grant No. NMP3-SL-2009-228989 ``OxIDes''. I thankfully 
 acknowledge the computer resources, 
 technical expertise and assistance provided by the 
 Red Espa\~nola de Supercomputaci\'on (RES) and by the 
Supercomputing Center of Galicia (CESGA).

\bibliography{max-feb28}

\end{document}